\documentclass[fleqn,usenatbib]{mnras}
\usepackage{newtxtext,newtxmath}
\usepackage{lipsum}

\usepackage[T1]{fontenc}
\usepackage{ae,aecompl}

\usepackage{graphicx}	% Including figure files
\usepackage{amsmath}	% Advanced maths commands
\usepackage{amssymb}	% Extra maths symbols

\newcommand{\dm}{$\Delta m_{15} (B)$}

\def\lsim{\hbox{\rlap{\raise 0.425ex\hbox{$<$}}\lower 0.65ex\hbox{$\sim$}}}
\def\gsim{\hbox{\rlap{\raise 0.425ex\hbox{$>$}}\lower 0.65ex\hbox{$\sim$}}}

\defcitealias{siebert19}{S19}
\defcitealias{fk11}{FK11}
\defcitealias{fsk11}{FSK11}
\defcitealias{Fitzpatrick}{Fitzpatrick}
\def\code#1{\texttt{#1}}

\title[Velocity May Improve SN~Ia Distances]{A Possible Distance Bias for Type Ia Supernovae with Different Ejecta Velocities}

\author[M. R. Siebert et al.]{
M. R. Siebert,$^{1}$\thanks{E-mail: msiebert@ucsc.edu}
R. J. Foley,$^{1}$
D. O. Jones,$^{1}$
K. W. Davis$^{1,2}$
\\
$^{1}$Department of Astronomy and Astrophysics, University of California, Santa Cruz, CA 95064\\
$^{2}$Department of Astronomy and Astrophysics, University of California Los Angeles, Los Angeles, CA 90095\\
}

\date{Accepted XXX. Received YYY; in original form ZZZ}

\pubyear{2019}

\begin{document}
\label{firstpage}
\pagerange{\pageref{firstpage}--\pageref{lastpage}}
\maketitle

\begin{abstract}
After correcting for their light-curve shape and color, Type Ia supernovae (SNe~Ia) are precise cosmological distance indicators.  However, there remains a non-zero intrinsic scatter in the differences between measured distance and that inferred from a cosmological model (i.e., Hubble residuals or HRs), indicating that SN~Ia distances can potentially be further improved. We use the open-source relational database \code{kaepora} to generate composite spectra with desired average properties of phase, light-curve shape, and HR. At many phases, the composite spectra from two subsamples with positive and negative average HRs are significantly different. In particular, in all spectra from 9~days before to 15~days after peak brightness, we find that SNe with negative HRs have, on average, higher ejecta velocities (as seen in nearly every optical spectral feature) than SNe with positive HRs. At +4 days relative to $B$-band maximum, using a sample of 62 SNe~Ia, we measure a $0.091\pm0.035$~mag ($2.7\sigma$) HR step between SNe with \ion{Si}{II} $\lambda$6355 line velocities ($v_{\ion{Si} {II}}$) higher/lower than $-11{,}000$~km~s$^{-1}$ (the median velocity). After light-curve shape and color correction, SNe with higher velocities tend to have underestimated distance moduli relative to a cosmological model. The intrinsic scatter in our sample reduces from 0.094~mag to 0.082~mag after making this correction.  Using the \ion{Si}{II} $\lambda$6355 velocity evolution of 115 SNe~Ia, we estimate that a velocity difference $>$500~km~s$^{-1}$ exists at each epoch between the positive-HR and negative-HR samples with 99.4\% confidence. Finally at epochs later than +37~days, we observe that negative-HR composite spectra tend to have weaker spectral features in comparison to positive-HR composite spectra.
\end{abstract}

% we measure negative-HR SNe~Ia to have an average \ion{Si}{II} $\lambda$6355 blueshift that is $1{,}000 \pm 200$ km s$^{-1}$ higher than positive-HR SNe~Ia. 
\begin{keywords}
supernovae: general
\end{keywords}

\section{Introduction}\label{sec:intro}

%-Type Ias
%-Cosmological distance indicators
%-HRs, unknown physics, corrections, mass step
%-Studies on HRs, environmental dependence
%-Studies on velocity (HVG, classes, color, etc.)
%-large sample for investigation

Type Ia supernovae (SNe~Ia) are important tools for understanding the acceleration of the expansion of the Universe caused by dark energy \citep{Riess98, Perlmutter99}. SN~Ia light-curve shape and color correlate with luminosity and a time series of photometric measurements of SNe~Ia allow us to measure precise distances to these phenomena \citep{pskovskii11, phillips93, riess96}. As cosmological surveys continue to reduce their calibration uncertainty, systematic uncertainties related to the explosion physics of SNe~Ia must also decrease or they will limit our ability to constrain dark energy properties \citep{jones18}. 

While light-curve shape and color are adequate to reduce the scatter in SN~Ia distances to $\sim$8\% \citep{scolnic18, Jones19}, more parameters may be needed to further improve precision. Hubble residuals (HRs) are the differences between the measured distance moduli and distance moduli inferred from a cosmological model. After making corrections for light-curve shape and color, the HR scatter cannot be explained by photometric uncertainty alone, and this {\it intrinsic} scatter could be related to progenitor environmental properties or explosion physics \citep{conley11, scolnic18}. Several studies \citep[e.g.,][]{kelly10,lampeitl10,sullivan10} have all found that the HR of a SN~Ia is correlated with its host-galaxy mass. Other studies have found that host-galaxy metallicity also correlate with corrected SN~Ia luminosity \citep{dandrea11, childress13, pan14}. These measurements should be a proxy for a physical property of the progenitor system. Current cosmological analyses treat the host-galaxy mass-HR relationship as a step function, yet we do not fully understand its origin or functional form \citep{childress13b}. 

\citet{rigault13} first studied how SN properties correlated with local host-galaxy properties and found evidence for a HR step with local host-galaxy star formation rate (SFR). \citet{rigault18} found reduced evidence for a local SFR step after using an updated version of the SALT2 SN\,Ia model \citep{Guy10,Betoule14}, but found strong evidence for a local {\it specific} SFR (sSFR) step, and \citet{roman18} found similar evidence for a local step by measuring host-galaxy $U-V$ colors. \citet{jones18} investigated the stellar masses and $u-g$ colors within 1.5~kpc of SN~Ia explosion sites, finding evidence for a local mass step, but they could not definitively say that local properties are better correlated with SN~Ia HRs than global properties or random information.  \citet{rose19} found a correlation between HR and global host-galaxy age, but did not see evidence for a stronger local effect.

The UV spectra of SNe~Ia could be a more direct probe of progenitor metallicity \citep[e.g.,][]{Hoflich98, Lentz00, foley08b, Sauer08, Walker12}. \citet{fk13} found that two ``twin" SNe~Ia (SNe~2011by and 2011fe) have very similar light curves, colors, and spectra, but different UV continua. They attribute this difference to a difference in progenitor metallicity. \citet{foley19} then showed that these SNe have different intrinsic luminosities, indicating that progenitor metallicity could be related to intrinsic scatter. Similar results were found for a larger, but more diverse sample \citep{Foley16}. \citet{pan19} looked at a larger sample of SN~Ia UV spectra and found that SN~Ia HRs are positively correlated with progenitor metallicity; however, \citet{brown19} did not find evidence for this UV flux-metallicity correlation.  Currently, spectral properties are infrequently used to calibrate SNe~Ia.

Numerous studies have quantified the potential for spectroscopic measurements to improve HR intrinsic scatter. \citet{bailey09} found that measuring flux ratios in specific wavelength bins could improve upon using light-curve parameters alone. \citet{blondin11} confirmed this result and investigated the relationships of several other spectral properties with HR. They found marginal evidence that measuring line velocities or absorption strengths improves HR scatter. \citet{silverman12} looked at an independent sample and found that velocity did not lead to significantly decreased HRs when applied in combination with SALT2 light-curve shape and color parameters. All of these studies found that flux ratios were the best spectral indicators for improving HR scatter. $\mathcal{R}(6520/4430)$, $\mathcal{R}^{c}(4610/4260)$, and $\mathcal{R}^{c}(3780/4580)$ \citep[respectively]{bailey09, blondin11,silverman12} were found to reduce HR scatter at >$2\sigma$ levels. Additionally, \citet{zheng18} modeled the relationship between peak magnitude, rise time, and photospheric velocity. They show that this model can significantly reduce HR scatter if high-velocity (HV) SNe are removed from the sample.

SNe~Ia have a large diversity of observed ejecta velocities \citep{branch87}. \citet{benetti05} found that a sample of SNe~Ia have a large range of \ion{Si}{II} $\lambda$6355 velocity gradients, where the amplitude of the gradient correlates with velocity at maximum light. \citet{wang09} classified ``High-Velocity'' and ``Normal'' SNe~Ia as SNe~Ia above and below a velocity of $\sim$11,800~km~s$^{-1}$ respectively. For typical SNe~Ia, the velocities derived from \ion{Si}{II} $\lambda 6355$ do not correlate with the decline-rate parameter \dm \space \citep{hatano00}. Some have suggested that ``High-Velocity'' SNe  come from different progenitor channels than ``Normal'' SNe \citep[e.g.,][]{Foley12:csm, maguire13, xiaofeng13}. Additionally, \citet{fk11} (hereafter \citetalias{fk11}) found that HV SNe~Ia have redder intrinsic $B-V$ colors than Normal SNe~Ia. \citetalias{fk11}  suggested that this effect could be reproduced with an asymmetric explosion viewed at different angles.  Using a sample of 1630 optical spectra of 255 SNe, \citet{fsk11} (hereafter \citetalias{fsk11}) measured a correlation between maximum-light ejecta velocity and intrinsic $B-V$ color (the velocity-color relationship; VCR). This result was also verified for high-redshift SNe~Ia \citep{Foley12:highz}.  If there are intrinsic differences between SNe~Ia with varying velocities, we might expect that our cosmological distance corrections do not fully account for these effects.

In this work, we aim to investigate whether any optical spectral properties of SNe~Ia correlate with HRs. This work makes use of \code{kaepora}, a public, open-source relational database of SN~Ia spectra that was recently presented by \citet{siebert19} (hereafter \citetalias{siebert19}). This tool provides a large sample of homogenized SN~Ia spectra and their associated metadata. For this analysis, we have updated \code{kaepora} with new metadata from the output of SALT2 fits to SN light curves \citep{jones18}. This new version of \code{kaepora} is available online\footnote{\href{https://msiebert1.github.io/kaepora/}{https://msiebert1.github.io/kaepora/}}. Also included is a set of tools that is useful for spectroscopic analysis. In this work we primarily use composite spectra to investigate SNe~Ia with varying HRs and control for spectral variation with phase and light-curve shape. These composite spectra cover a large wavelength range, and can provide more information about potential spectral feature trends than individual measurements (like ejecta velocities or equivalent widths). 

In Section~\ref{sec:methods}, we describe the subsample of SNe in \code{kaepora} that have HR measurements which we then use for this study. In Section~\ref{sec:results}, we present our HR-binned composite spectra and measure trends of spectral features with HRs. In Section~\ref{sec:disc}, we summarize these spectral trends and discuss how they might impact cosmology. We conclude in Section~\ref{sec:conc}.

\section{Methods} \label{sec:methods}

\subsection{Sample}
Version 1.1 of \code{kaepora} contains 4975 spectra of 777 SNe~Ia. The majority of these data are sourced from the Center for Astrophysics (CfA) Supernova Program \citep{CfA}, the Berkeley SN Ia Program (BSNIP; \citealt{bsnip}), and the Carnegie Supernova Project (CSP; \citealt{CSP}) accounting for 51.9\%, 25.6\%, and 12.4\% of the total spectra, respectively. While the CfA sample contains many spectra per SN (typically 7), the BSNIP spectra cover a much larger wavelength range (typical ranges of $\sim$3500 -- 7500~\AA\ and $\sim$3200 -- 10000~\AA, respectively). We also include data from \citet{gomez96, riess97, leonard05, blondin06, matheson08, Foley12:csm} and \citet{silverman15}. This relational database allows for these data to be associated with a variety of useful SN-specific and spectrum-specific metadata. 

\begin{figure}
    \includegraphics[width=3.2in]{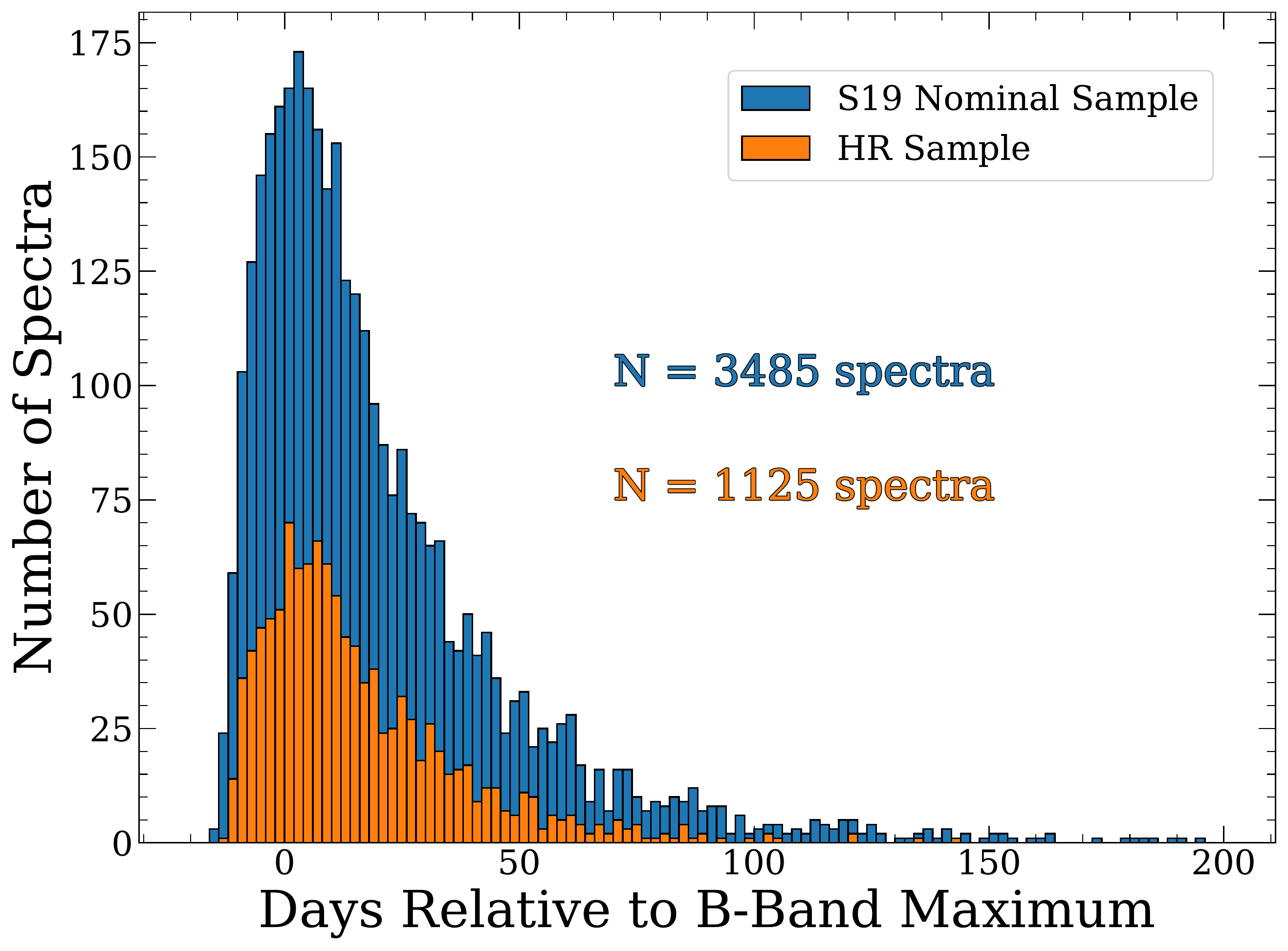}
  \caption{Blue and orange histograms of the number of individual spectra per SN in the \citetalias{siebert19} nominal and HR samples, respectively. The median phases are +12 and +10 days, respectively.}\label{fig:dem1}
\end{figure}

\begin{figure}
    \includegraphics[width=3.2in]{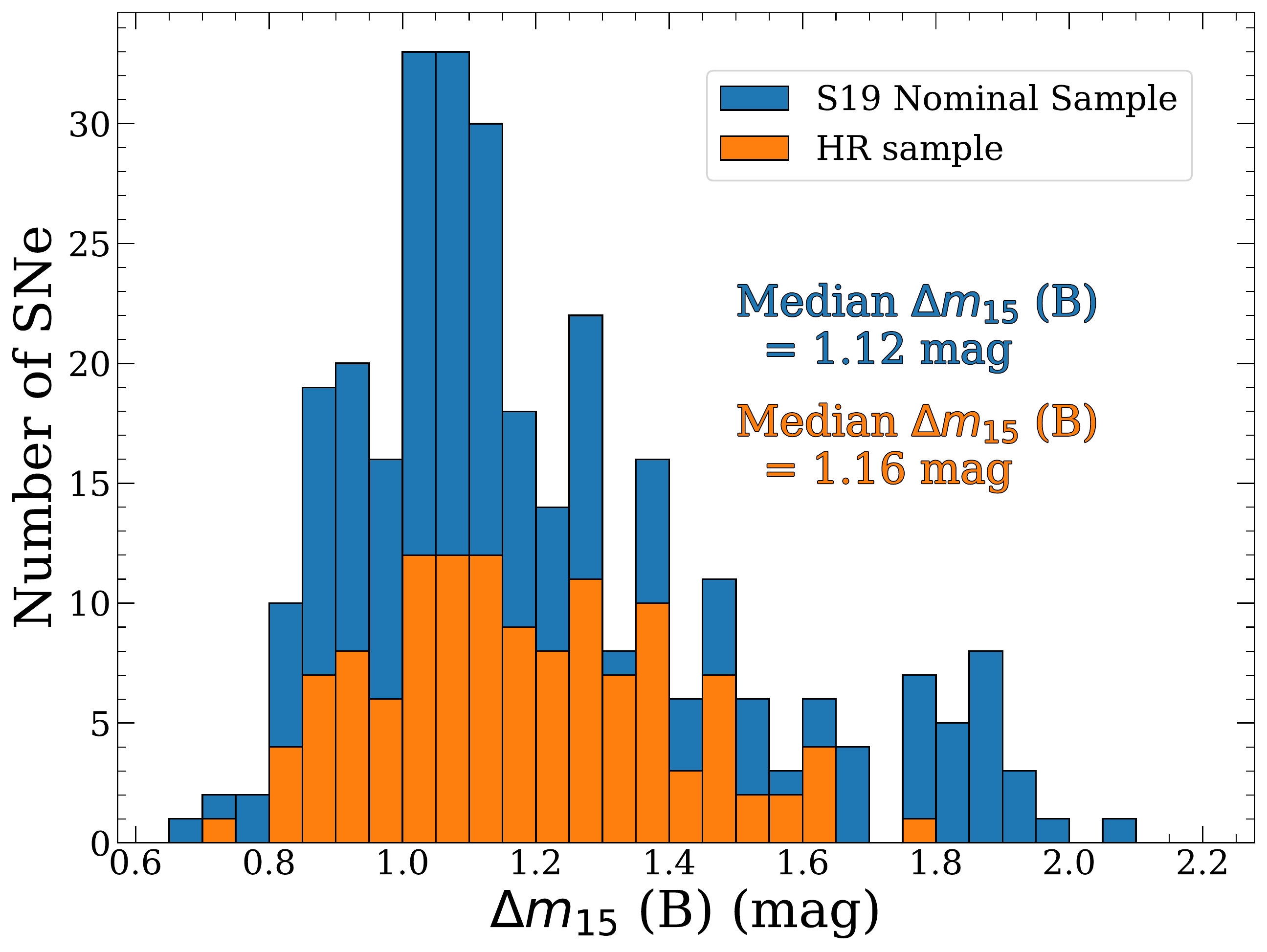}
  \caption{Blue and orange histograms of the number of SNe~Ia per \dm\ bin in the \citetalias{siebert19} nominal and HR samples, respectively.}\label{fig:dem2}
\end{figure}

\begin{figure}
    \includegraphics[width=3.2in]{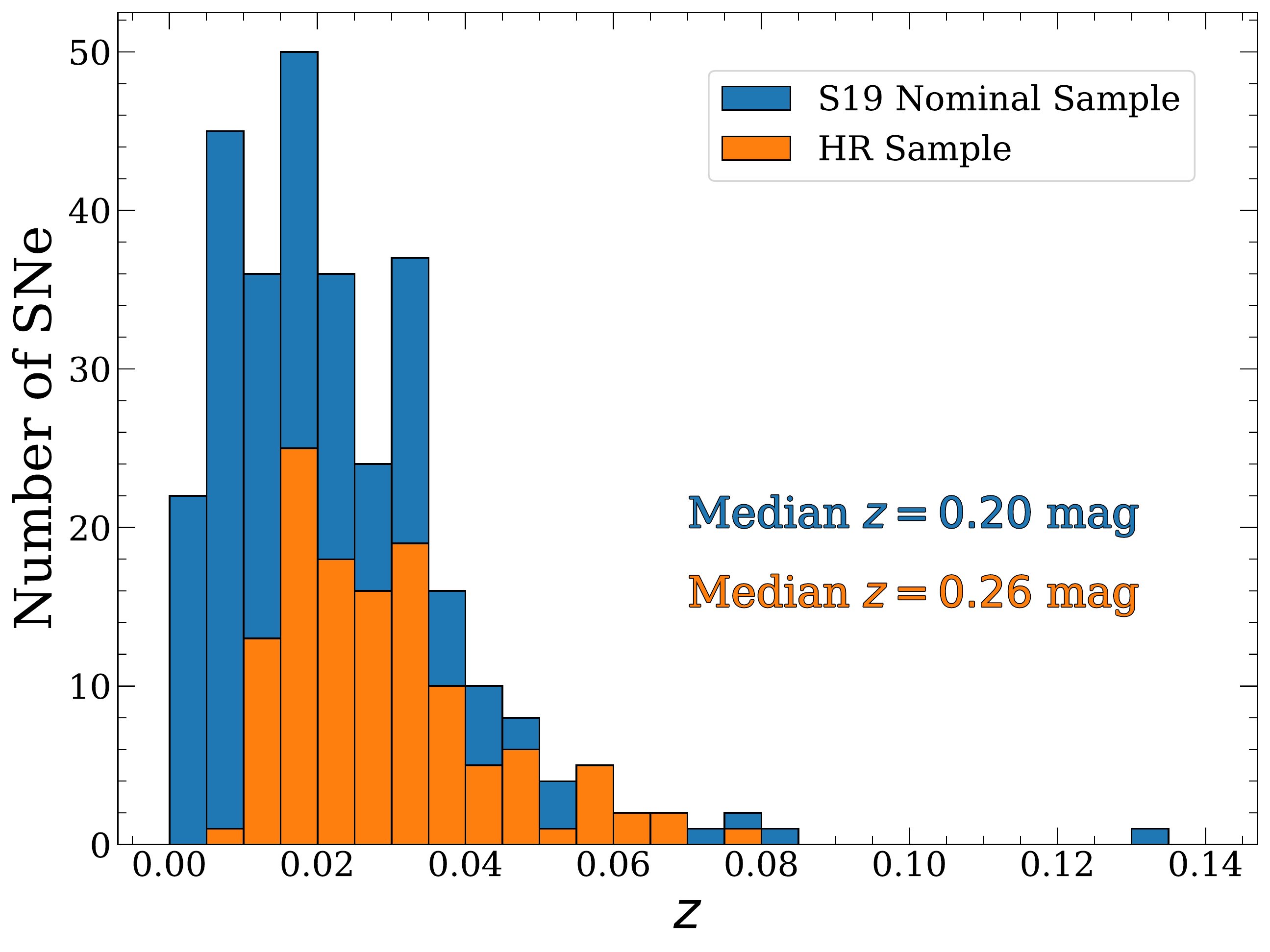}
  \caption{Same as Figure~\ref{fig:dem2}, but for redshift.}\label{fig:dem3}
\end{figure}

\begin{figure}
    \includegraphics[width=3.2in]{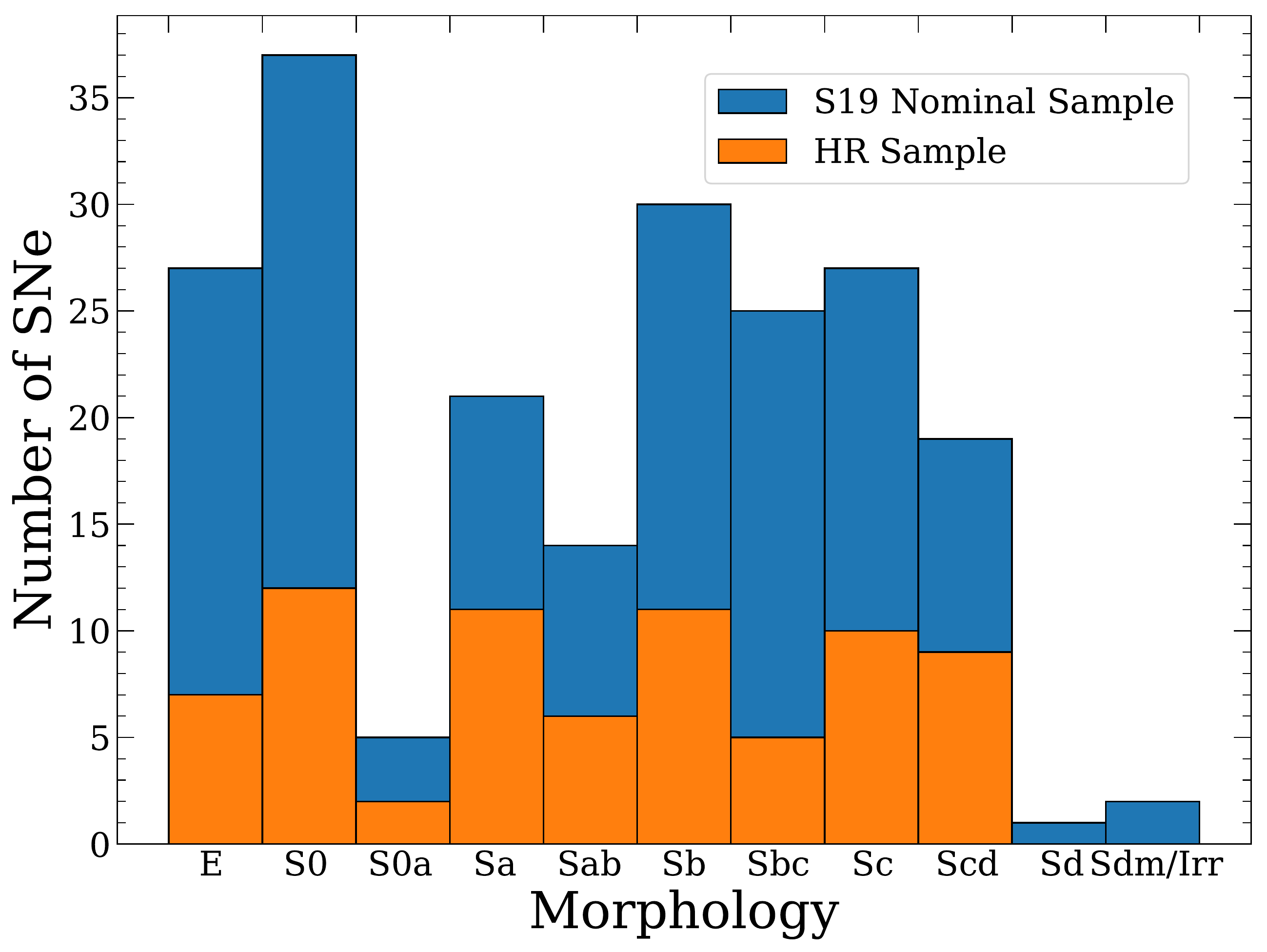}
  \caption{Same as Figure~\ref{fig:dem2}, but for host-galaxy morphology.}\label{fig:dem4}
\end{figure}

The spectral sample for this work is a subset of the ``nominal" sample described in detail by \citetalias{siebert19}. The nominal sample required SNe with a calibrated light curve that covered the time of maximum light and an MLCS2k2 \citep{jha07} host-galaxy extinction estimate in order to correct the spectra for host-galaxy dust reddening. The MLCS2k2 light-curve fits adopt $R_{V} = 2.5$ since typical cosmological analyses often exclude highly reddened SNe~Ia that seem to favor lower values of $R_{V}$ \citep{kessler09, burns14}.  We also correct each spectrum for Milky Way (MW) reddening using the \citet{S&F} reddening map, a \citet{Fitzpatrick} reddening law, and $R_{V} = 3.1$.

\subsection{Hubble Residual Measurements}
The HRs used in this analysis are from \citet{Jones18b}, with the SN light curves themselves originating from CfA surveys 1-4 \citep{riess99,Jha06,hicken09,hicken12} and CSP \citep{contreras10,folatelli10,stritzinger11}, with additional SNe from SDSS \citep{kessler09} and Pan-STARRS \citep{Rest14,jones18,scolnic18} that are not included in this work.  Notably, Foundation Supernova Survey light curves \citep{Foley18} are not included in this analysis, but will be in a future paper (Dettman et~al., in prep.).

Light-curve fits to these data used the most recent version of the SALT2 model, SALT2.4 \citep{Guy10,Betoule14}.  We also applied the selection criteria used in the Pantheon analysis \citep{scolnic18} to ensure accurate distances.  These include shape and color cuts that remove SNe with parameters outside the range for which the model is valid ($−3 < x_{1} < 3$, $−0.3 < c < 0.3$) and cuts on the uncertainties of the shape and time of maximum light.  We require MW reddening of $E(B-V) < 0.15$~mag and $z > 0.01$ to remove SNe that have distances with large peculiar-velocity uncertainties.  The color cuts avoid SNe~Ia with large dust reddening, which has been linked with atypical dust reddening and possible circumstellar scattering \citep[e.g.,][]{Elias-Rosa06, Wang08:06x, Phillips13, Foley14, Goobar14} and has been linked to spectral differences \citep{wang09, Mandel14}.

SN~Ia distances are derived from the light-curve parameters using a version of the \citet{Tripp98} formula that includes the host-galaxy mass step, $\gamma$ and a bias correction, $\Delta_B(z)$:
\begin{equation}
  \mu = m_B - \mathcal{M}\ + \alpha \times x_1 - \beta \times c + \Delta_B(z) + \gamma.
  \label{eqn:salt2}
\end{equation}
Here, $m_B$ is the light curve amplitude, $\mathcal{M}$ is the SN absolute magnitude (whose exact value is degenerate with that of the Hubble constant and is irrelevant when only comparing HRs), and $\alpha$ and $\beta$ are nuisance parameters that are determined from a fit to the full low-$z$ sample.
This fit is performed using the BEAMS with Bias Corrections (BBC) method of \citep{Kessler17}, which corrects for observational biases on the $x_{1}$, $c$, and $m_{B}$ parameters as well as on $\alpha$ and $\beta$.  Those biases are estimated using large Monte Carlo simulations, generated with the SNANA software \citep{Kessler09b}, to match the low-$z$ samples observations and sample demographics.  Biases are corrected with the $\Delta_B(z)$ term.  We estimate HRs relative to the maximum likelihood distances in three redshift bins to remove any dependence on cosmological parameters.

\subsection{Hubble Residual Spectroscopic Sample}
This work investigates the intersection of the \citetalias{siebert19} nominal sample with the sample of SNe~Ia for which we have also measured HRs (named the HR sample). These requirements limit the HR sample to 1125 spectra of 126 SNe~Ia. The properties of this subsample are compared to those of the \citetalias{siebert19} nominal sample in Figures~\ref{fig:dem1} -- \ref{fig:dem4}. Overall, the distributions of the parameters of the HR sample is similar to the \citetalias{siebert19} nominal sample. Since cosmological analyses cut out the fastest-declining SNe, the HR sample does not cover as wide a range of \dm\ as \citetalias{siebert19}. The HR sample also has a slightly larger median redshift ($z = 0.026$ in the HR sample versus $z = 0.020$ in \citetalias{siebert19}); this is because cosmological analyses only include SNe~Ia in the Hubble flow where peculiar velocities of the SN host galaxies have a sufficiently small impact on distance measurements. It is also important to note that the HR sample only includes 7 spectra with $\tau > +100$~days compared to 131 spectra in the \citetalias{siebert19} nominal sample. Thus, analyses of the spectra in the HR sample at late phases are difficult to perform.

% \begin{figure}
%     \includegraphics[width=3.2in]{HR_demo_av_dist.pdf}
%   \caption{Histogram of the number of SNe~Ia per host-galaxy extinction bin.}\label{fig:dem5}
% \end{figure}

\section{Analysis}\label{sec:results}
The general goal of this work is take an agnostic approach to investigating the spectral properties of SNe~Ia that could be related to HRs determined with current photometric-only techniques. Several of the studies discussed in Section~\ref{sec:intro} first form a hypothesis about how certain observables (e.g., host SFR, progenitor metallicity, etc.) may correlate with HRs, then focus on those individual measurements. Here, we take an alternative approach where we assume that the intrinsic scatter is caused by physical differences that can manifest as spectral differences. This approach minimizes the bias associated with determining the importance of specific pre-determined observables, can more easily discover unexpected results, and if no differences are found, produces a limit on the importance of spectral variability for reducing potential distance biases. \code{kaepora}, which both has tools to produce composite spectra for subsets of SNe and can easily control for properties known to correlate with the property in question (i.e., HR in this case), is uniquely designed for this kind of analysis. By looking at the average spectra of SNe with different HRs, we aim to determine which spectral properties (if any) correlate with HR.

\begin{figure*}
\begin{center}
    \includegraphics[width=6.1in]{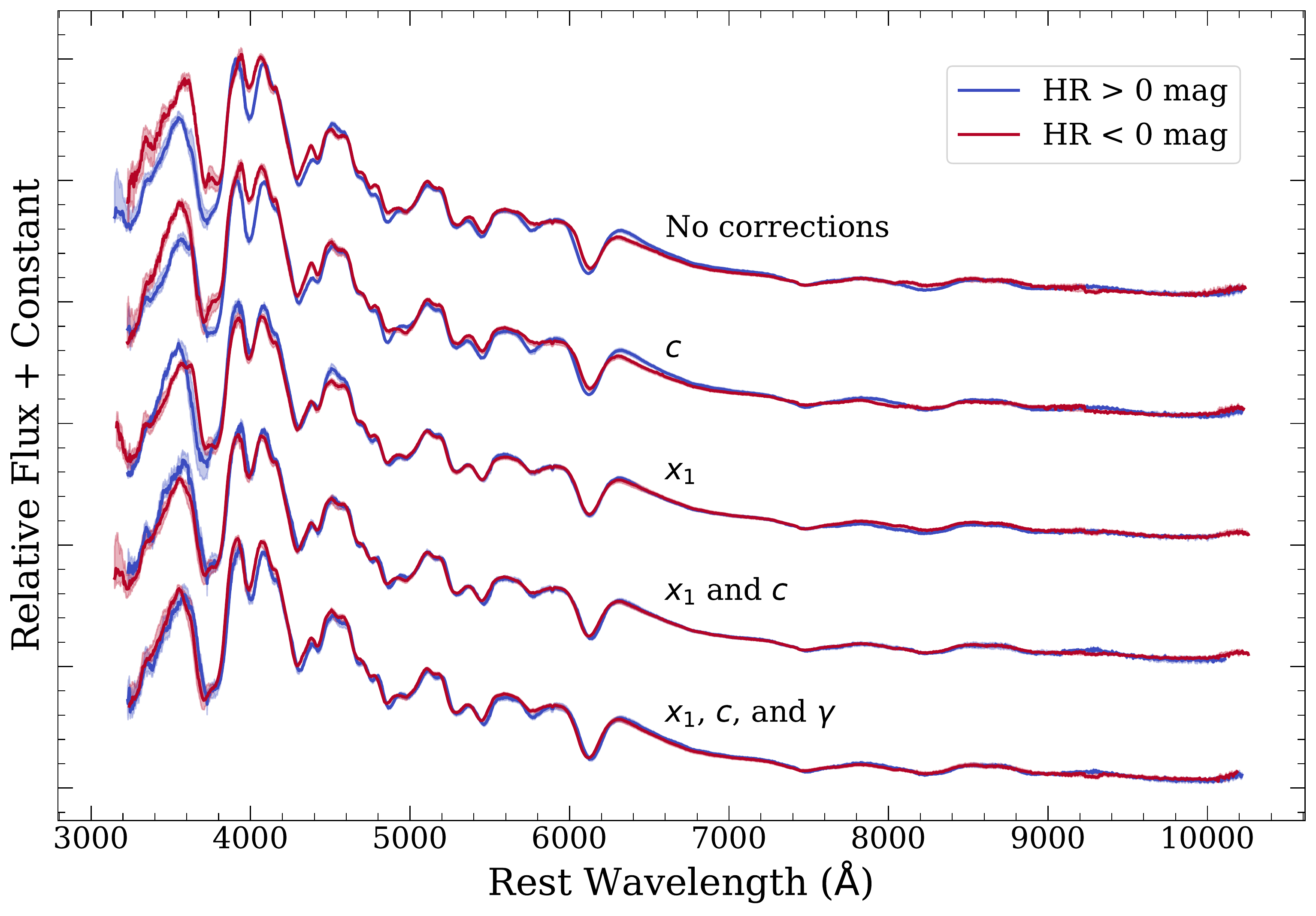}
\caption{Four sets of maximum-light ($-2 < \tau < +2$~days) composite spectra constructed using positive- (blue) and negative-HR (red) bins. The shaded regions are the 1-$\sigma$ bootstrap-sampling uncertainty for each composite spectrum. From top to bottom, the composite spectra vary in the distance modulus corrections that have been applied before measuring HRs. We start by using uncorrected luminosities, then subsequently correct for only light-curve shape ($x_1$); only color ($c$), light-curve shape and color ($c$); and light-curve shape, color, and host-galaxy mass step ($\gamma$). Most of the spectral variation between these two samples is removed after making the $x_1$ and $c$ corrections.}\label{fig:hr_corr}
\end{center}
\end{figure*}

\subsection{Composite Spectra}
In this work we generate Gini-weighted composite spectra using the methods presented by \citetalias{siebert19} for a variety of subsets of the HR sample. The Gini-weighting method provides a representative spectrum that maximizes the signal-to-noise ratio (S/N) while reducing the impact of individual high-S/N outliers. The basic algorithm is adapted from the methods used by \citet{foley08} and \citet{foley12:sdss}. The composite spectra also contain information about the uncertainty and average properties (e.g., phase, \dm, and redshift) as a function of wavelength. This information allows us to control for SN parameters that have known correlations with spectral features (such as phase and light-curve shape) as well as potentially unknown biases. \citetalias{siebert19} showed that a subset of spectra is well represented by a composite spectrum with the same average properties. We implement the same bootstrap resampling with replacement as in \citetalias{siebert19} in order to estimate the sample variation about the average spectrum.

In this work we use \code{kaepora} to provide subsets of SN~Ia spectra that have similar average values of phase and \dm\ but different average HRs. By controlling for the main parameters that influence the diversity of SN~Ia spectra, the remaining differences between composite spectra should be related to the parameter being varied. In this Section we present many of these Gini-weighted composite spectra. Due to the limitations of our sample size, we primarily present composite spectra generated using two HR bins (positive and negative). In \autoref{fig:hr_corr} we present five sets of maximum-light ($-2 < \tau < +2$~days) composite spectra constructed using these HR bins. From top to bottom, the composite spectra vary by the number of corrections that are made to the distance moduli of the contributing SNe. 

First, we examine the composite spectra generated from SNe with positive and negative HRs when no light-curve shape, color, or mass-step corrections are applied.  In this particular case, a SN with a negative HR has a higher peak luminosity and/or less host-galaxy dust extinction than a SN with a positive HR.  The effective HR of these composite spectra are $0.22$ and $-0.24$~mag, respectively.  For these composite spectra, we see spectral differences between the positive-HR and negative-HR (higher peak luminosity) composite spectra that correspond to known trends in spectral features with light-curve shape. For example, the Si ratio ($\mathcal{R}$(\ion{Si}{II}); \citealt{nugent}) is smaller in the negative-HR composite spectrum.  Additionally, the overall continua of the negative-HR composite spectrum is bluer than the positive-HR composite spectrum, consistent with intrinsic colors of different luminosity SNe~Ia. These results confirm that our method can reproduce expected spectral differences.

Applying only a color correction (i.e., correcting $\mu$ by $c\beta$) provides the biggest reduction in HR scatter for a single correction. The positive-HR and negative-HR composite spectra have effective HRs of  $0.03$ and $-0.10$~mag, respectively. Applying only a light-curve shape correction (i.e., correcting $\mu$ by $\alpha x_{1}$) decreases the spectral differences of the positive-HR and negative-HR samples, but the HR difference does not decrease as significantly. The effective HR of the composite spectra for these subsamples are $0.24$ and $-0.22$~mag, respectively. The spectral features of these composite spectra are well matched and the most notable difference is in the continua. The negative-HR composite spectrum (red curve labeled ``$x_1$ correction") appears to be slightly redder in color than its positive-HR counterpart. Naively one might expect the opposite relationship (the distances of bluer SNe after light-curve shape correction should be underestimated if the difference corresponds only to a lack of a dust-reddening correction). However, if the $x_1$ correction also accounts for continuum differences in some way, then it is possible to have this relationship with a color correction necessary to match continua. There is also a slight difference in the Ca H\&K and Ca NIR triplet features. The Ca features in the negative-HR composite spectrum are slightly weaker and at lower velocity than the positive-HR composite spectrum.

Correcting distance moduli for light-curve shape and color (correcting $\mu$ for both $\alpha x_1$ and $c \beta$) yields HR-binned composite spectra that are very similar. The continua are almost identical and the spectral feature strengths are well matched. The effective HR of these composite spectra are $0.10$ and $-0.12$~mag, respectively, indicating that applying these distance modulus corrections has significantly reduced the HR scatter as expected. 

Finally, introducing an additional host-mass step correction ($\gamma$) produces little change to the spectral features of the HR-binned composite spectra compared to those produced without making the host-mass correction. The effective HR of these composite spectra are $0.10$ and $-0.09$~mag, respectively. 

For the remainder of this work, composite spectra are generated using distance moduli that have been corrected for $x_1$, $c$, and $\gamma$, corresponding to the values used in cosmological analyses, unless otherwise noted. 

\subsection{Hubble Residuals and Velocity}

\begin{figure}
\begin{center}
\includegraphics[width=3.2in]{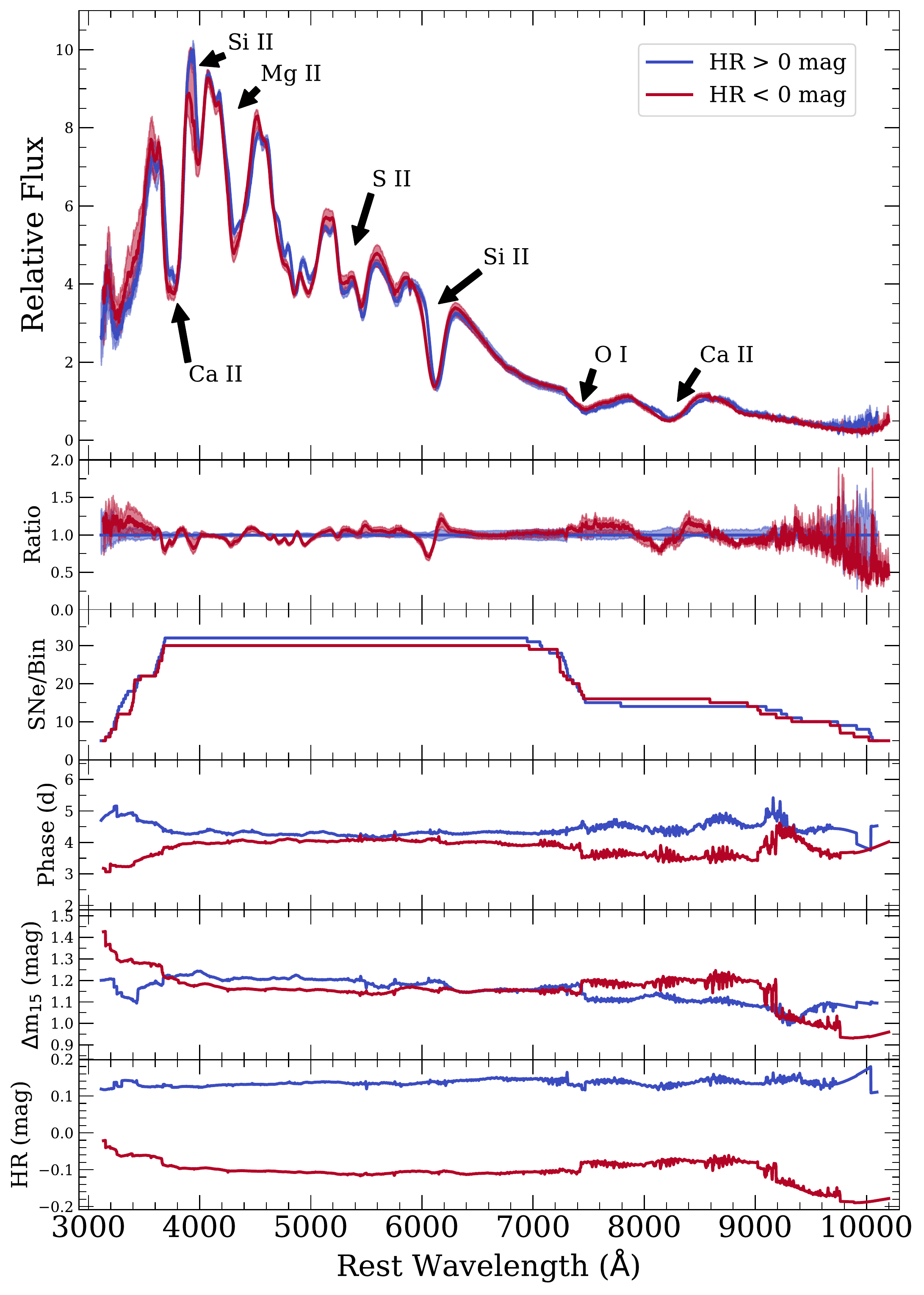}
\caption{(\textit{First panel}): +4-day composite spectra created from our nominal sample ($+2 < \tau < +7$~days, $0.7 \leq$ \dm\ $\leq 1.8$~mag). The blue curves show the properties of our positive-HR composite spectrum and the red curves show the properties of our negative-HR composite spectrum. The shaded regions are the 1-$\sigma$ bootstrap sampling uncertainties. (\textit{Second panel}): Ratio of the negative-HR composite spectrum relative to the positive-HR composite spectrum (red) and uncertainties on the positive-HR/negative-HR composite spectra (blue/red). (\textit{Third panel}): Number of individual spectra contributing to each wavelength bin. (\textit{Fourth panel}): Average phase relative to maximum brightness as a function of wavelength. (\textit{Fifth panel}): Average value of \dm\ as a function of wavelength. (\textit{Sixth panel}): The average HR as a function of wavelength.}\label{fig:p4_panels}
\end{center}
\end{figure}

\begin{figure*}
    \includegraphics[width=6.4in]{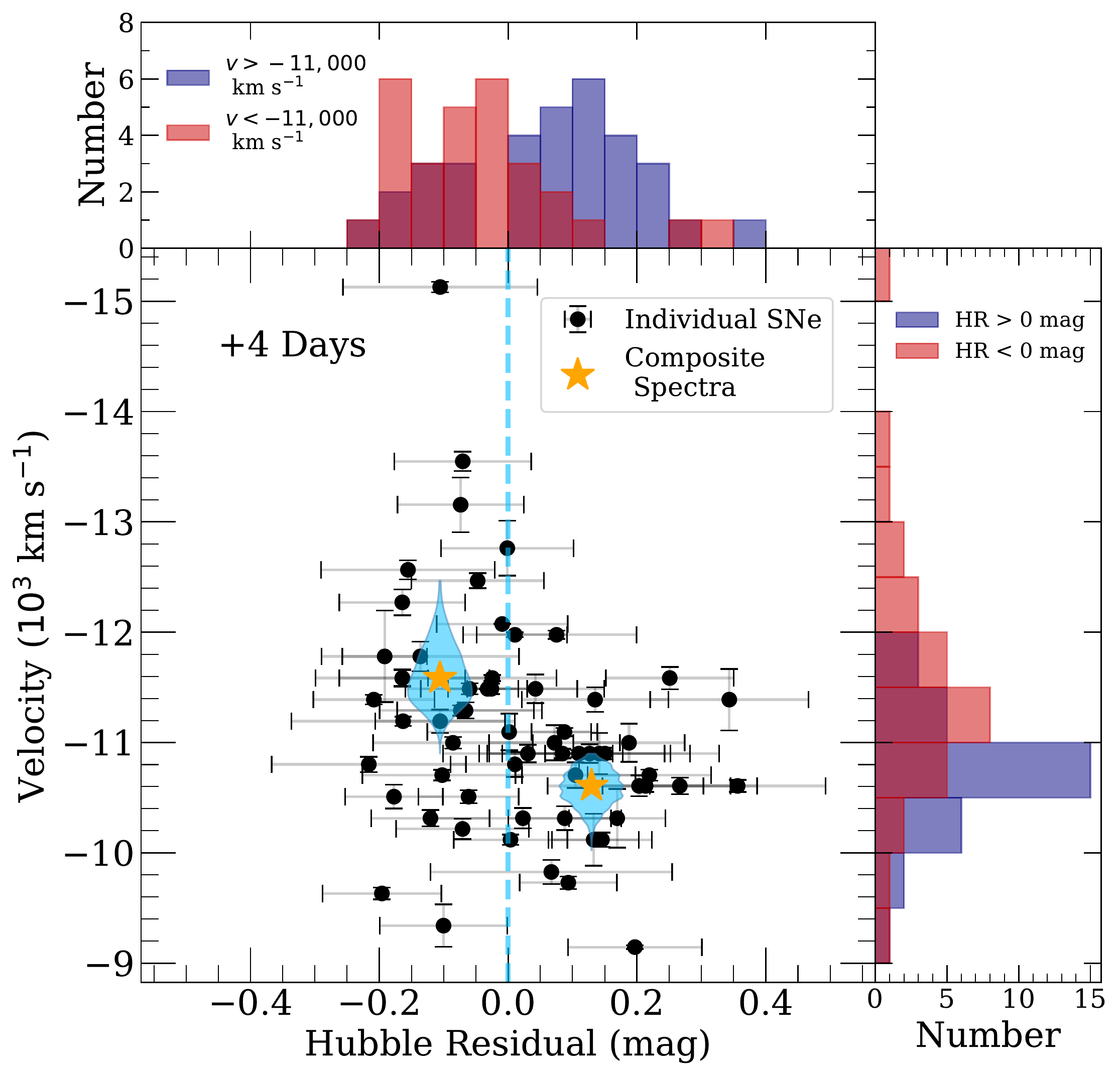}
  \caption{\ion{Si}{II} velocity for individual SNe versus their HR (black points) for the sample of SNe contributing to our +4-day composite spectra in \autoref{fig:p4_panels}.  Orange stars correspond to the measurements from our composite spectra. The blue shaded region shows the distribution of velocities from bootstrap resampling. The vertical blue-dashed line at ${\rm HR} = 0$~mag shows where the sample is divided. The top blue and red histograms display the distributions of HRs for SNe with $v > -11,000$~km~s$^{-1}$ and $v < -11,000$~km~s$^{-1}$, respectively. The right blue and red histograms display the distributions of velocities for positive-HR and negative-HR SNe, respectively.}\label{fig:p4_vs_all}
\end{figure*}

\begin{figure}
\begin{center}
\includegraphics[width=3.2in]{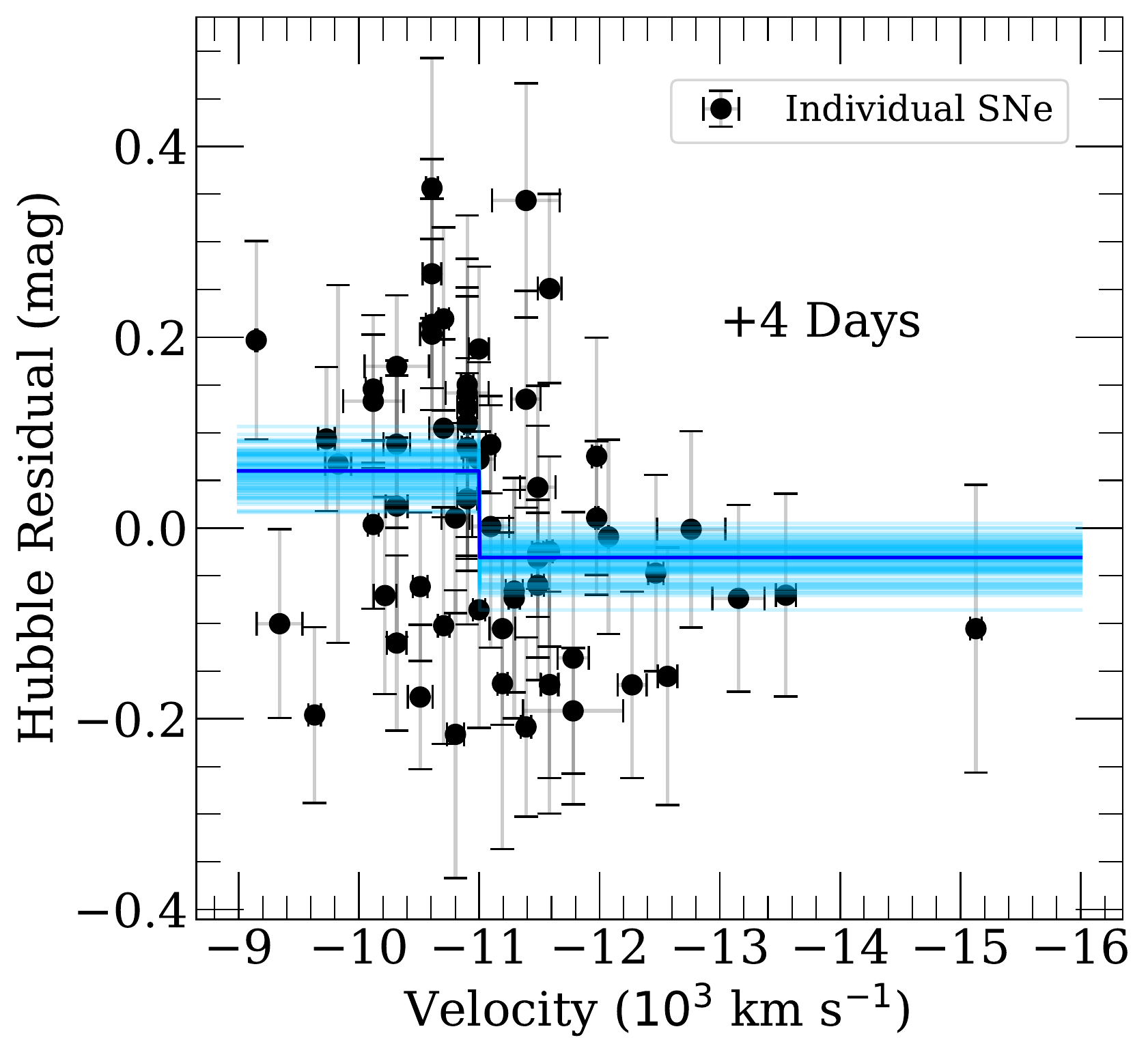}
\caption{HR versus \ion{Si}{II} velocity for individual SNe (black) in the +4-day sample. We fit a two-parameter step function (blue) to these data where we split the sample at the median velocity ($-11{,}000$~km~s$^{-1}$). The light-blue lines represent a subset of fits to random realizations of the data. The orange stars show measurements from composite spectra where we separate the sample at the same velocity. The best-fitting offset between two subsamples is $0.091 \pm 0.035$~mag offset ($2.7\sigma$).}\label{fig:hrstep}
\end{center}
\end{figure}

% \begin{figure}
% \begin{center}
% \includegraphics[width=3.2in]{hr_step.pdf}
% \caption{}\label{fig:hrstep}
% \end{center}
% \end{figure}

Despite having similar continua and overall line strengths, we observe a general trend that composite spectra generated from SNe with differing HRs tend to have different line velocities in many of their spectral features. In \autoref{fig:p4_panels}, we present composite spectra generated from samples of SNe with positive and negative HRs (blue curve) and spectra with phases of $+2 < \tau < +7$~days.  The positive-HR and negative-HR composite spectra were constructed from 88 spectra of 30 SNe, and 65 spectra of 32 SNe, respectively.  The composite spectra have effective phases of +4.4 and +3.8~days and effective values of \dm\ of 1.13 and 1.19~mag, respectively.
%In the first panel of \autoref{fig:p4_panels}, we present two composite spectra, one generated from spectra of SNe with positive HRs (blue curve) and another generated from SNe with negative HRs (red curve). The second panel shows the ratio of the relative fluxes of the two composite spectra and the third panel shows the number of SNe contributing to a given wavelength bin.  Finally, the bottom three panels show the average properties of phase, \dm, and HR for the composite spectra as a function of wavelength, respectively. The positive-HR  and negative-HR composite spectra were constructed using a phase bin of $+2 < \tau < +7$ days and have effective phases of +4.4 and +3.8 days and effective values of \dm\ of 1.13 and 1.19 mag, respectively. Also, the positive-HR and negative-HR composite spectra were constructed from 88 spectra of 30 SNe, and 65 spectra of 32 SNe, respectively.

The positive-HR and negative-HR composite spectra have \ion{Si}{II} $\lambda$6355 line velocities of $-10{,}600$~km~s$^{-1}$ and $-11{,}580$~km~s$^{-1}$, respectively. Similar velocity shifts are also clearly visible for Ca H\&K, \ion{Si}{II} $\lambda$5972,6355, \ion{S}{II}, and the Ca near-infrared (NIR) triplet (see the second panel of \autoref{fig:p4_panels}). In regions where the continua are matched up well (e.g., \ion{Si}{II} and the Ca NIR triplet), these differences are obvious in the ratio of the composite spectra. 

Since the HR-binned composite spectra at +4~days show a dramatic difference in ejecta velocity, we also investigate the \ion{Si}{II} line velocities of the individual SNe contributing to each composite spectrum. The +4-day composite spectra contain data from 62 SNe whose velocity and HR measurements are shown in \autoref{fig:p4_vs_all} as black points (hereafter the ``+4-day" HR sample). The orange stars are the velocity measurements made directly from the composite spectra displayed in \autoref{fig:p4_panels} and the shaded-blue violin regions depict the distributions of velocities measured from the individual output spectra from our bootstrap resampling process of each sample.
%A blue-dotted vertical line was placed at HR = 0 mag to better depict which individual spectra contribute to each composite spectrum.
At the top of \autoref{fig:p4_vs_all}, we display HR histograms for the sample divided at $-11{,}000$~km~s$^{-1}$, the median velocity of the sample.
%($v_{Si\; \mathrm{II}} < -11{,}000$ km s$^{-1}$; blue, $v_{Si\; \mathrm{II}} > -11{,}000$ km s$^{-1}$; red).
We also display velocity histograms for positive-HR and negative-HR SNe on the right side of \autoref{fig:p4_vs_all}. 

There is no strong correlation between \ion{Si}{II} velocity at $+4$~days and HR (absolute Pearson correlation coefficient of $\sim$0.30), however there is some intriguing structure to the distribution of measurements. For example, there are no positive-HR SNe with velocities higher than $-12{,}000$~km~s$^{-1}$ in this phase bin. Using the uncertainties estimated from bootstrap resampling, we measure a velocity difference of $980 \pm 220$~km~s$^{-1}$ ($680 \pm 150$~km~s$^{-1}$ after removing $>${}$2\sigma$ velocity outliers) between the positive-HR and negative-HR composite spectra. We perform a Kolmogorov-Smirnov test with high/low-velocity ($v_{\mathrm{Si\; II}}$ below/above $-11{,}000$ km s$^{-1}$) samples and find a $p$-value of 0.00031 ($3.4\sigma$), suggesting that we can reject the hypothesis that the HR distributions of these samples are drawn from the same population. We also perform a Kolmogorov-Smirnov test with positive-HR/negative-HR samples and find a p-value of 0.00078 ($3.2\sigma$), suggesting that we can also reject the hypothesis that the velocity distributions of these samples are drawn from the same population. 

Despite the formally high significance of these tests, the HR uncertainties for individual objects are relatively large (typically $0.12$~mag), and a large fraction of SNe have a sizeable probability of truly belonging to the opposite group than it is assigned. To include this uncertainty in our significance tests, we generate $10^{6}$ realizations of the HR values in the +4-day sample, varying the HRs by a Gaussian distribution with a standard deviation corresponding to the uncertainty of the HR measurement. In \autoref{fig:hrstep}, we display HR as a function of \ion{Si}{II} velocity at +4~days after maximum light. The black points and error bars are the same data as in \autoref{fig:p4_vs_all} (but the axes are swapped). The dark-blue curve is a two-parameter step function fit to the original data where the sample is split at the median velocity ($-11{,}000$ km s$^{-1}$). The light-blue curves are 100 randomly chosen step-function fits to the resampled HR values. Using these data, we estimate a velocity step of $0.091 \pm 0.025$~mag ($3.7\sigma$). To investigate the importance of choosing the median velocity to separate the sample, we allowed the velocity that separates the samples to be a free parameter in our fit. Doing this, we measure a similar velocity step of $0.091 \pm 0.027$~mag ($3.4\sigma$), indicating that the exact choice of velocity to separate the sample is not driving the results. The removal of $>${}$2\sigma$ velocity outliers also has a small impact on the step size ($0.087 \pm 0.025$~mag). The low- and high-velocity samples have a sample intrinsic scatter of 0.108~mag and 0.064~mag, respectively. If we include the intrinsic scatter in the HR uncertainties, we measure a velocity step of $0.091 \pm 0.035$~mag ($2.7\sigma$).

\subsection{Temporal Evolution of the Velocity-HR Trend}
\begin{figure}
    \includegraphics[width=3.2in]{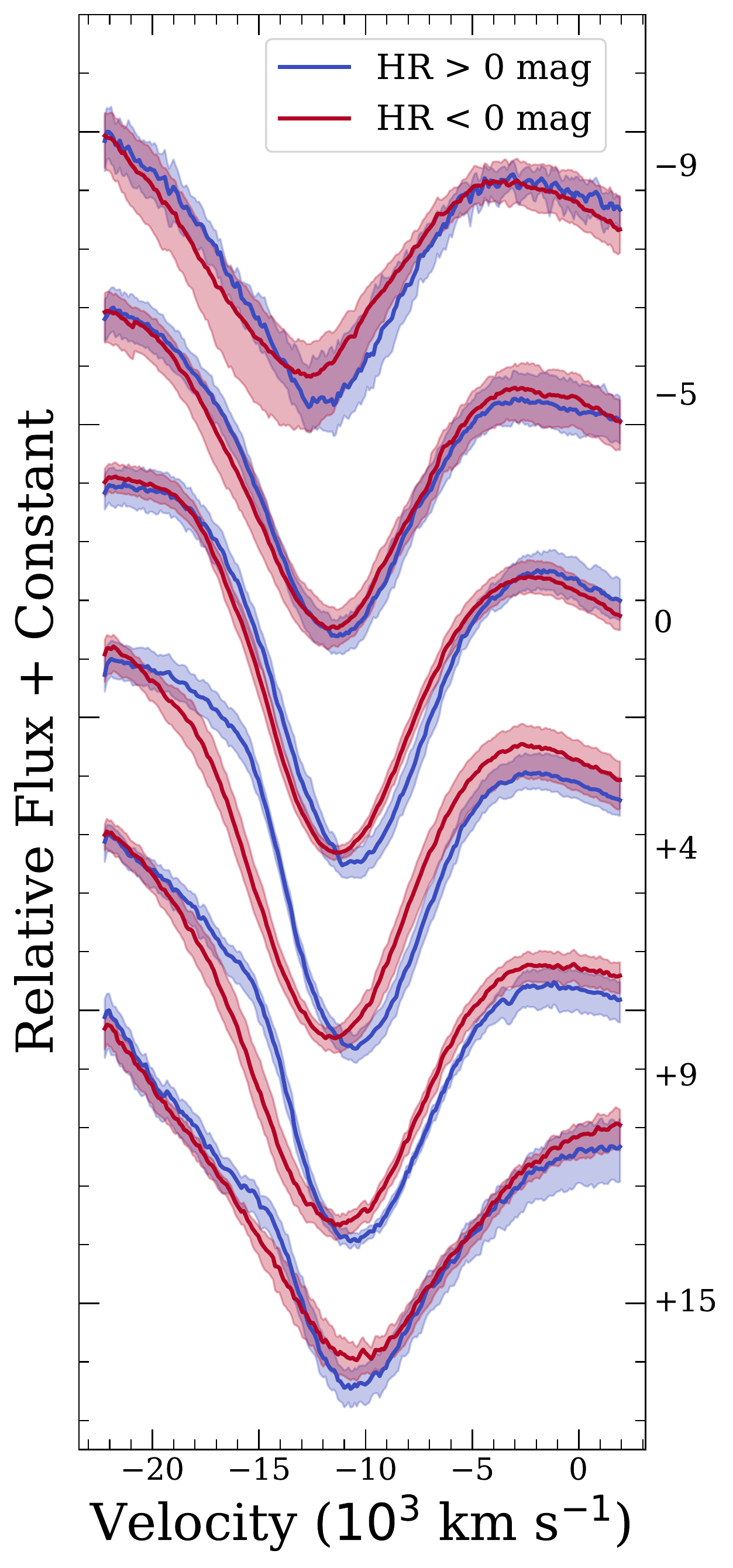}
  \caption{Time series of composite spectra created from subsamples of SNe with positive (blue) and (negative) HRs displaying the region around \ion{Si}{II} $\lambda$6355. The blue- and red-shaded regions are the 1-$\sigma$ bootstrap-sampling uncertainties of the positive- and negative-HR composite spectra, respectively. The right vertical axis indicates the effective phase in days of each set of composite spectra. At all epochs prior to +15~days, the minimum of the main \ion{Si}{II} feature is more blueshifted in the negative-HR sample than the positive-HR sample. }\label{fig:si_zoom}
\end{figure}

The velocity-HR trend is also visible over a large range of phases.  In \autoref{fig:si_zoom} we present the \ion{Si}{II} $\lambda$6355 feature in six positive-HR (blue curves) and six negative-HR (red curves) composite spectra representing phases of approximately $-9$ ($\tau < -7$), $-5$ ($-7 < \tau < -2$), $0$ ($-2 < \tau < +2$), $+4$ ($+2 < \tau < +7$), $+9$ ($+7 < \tau < +13$), and $+15$ ($+13 < \tau < +21$) days. At all epochs prior to $+15$~days after maximum light, we see that the \ion{Si}{II} feature appears more blueshifted in the negative-HR composite spectra (red) than the negative-HR composite spectra (blue). These differences are apparent across a large range of velocities and often can not be accounted for by the $1\sigma$ bootstrapping uncertainty regions.  The velocity difference manifests as a shift in the wavelength of maximum absorption (from which we determine the velocity of the feature), the width of the feature, and the position of the blue edge of the feature.  The full line profile provides additional information and evidence that the difference in ejecta velocity for the different HR subsamples is significant. 

Using the relationships derived by \citet{fsk11}, we use the spectrum closest to maximum light to estimate $v^0_{\mathrm{Si\; II}}$, the maximum-light \ion{Si}{II} velocity, for each SN. With a sample of 115 SNe~Ia we measure a $v^0_{\mathrm{Si\; II}}$-HR step of $0.068 \pm 0.027$~mag when the sample intrinsic scatter is included in the HR uncertainties. This is consistent with our step measurement from the +4-day sample ($0.091 \pm 0.035$ mag).

\subsection{Velocity-HR Trend Seen in Other Features}
The velocity-HR trend is also visible in several absorption features besides \ion{Si}{II} and over a large range of phases.  Here we examine these other indicators of ejecta velocity correlating with HR.

\begin{figure*}
\begin{center}
\includegraphics[width=6.1in]{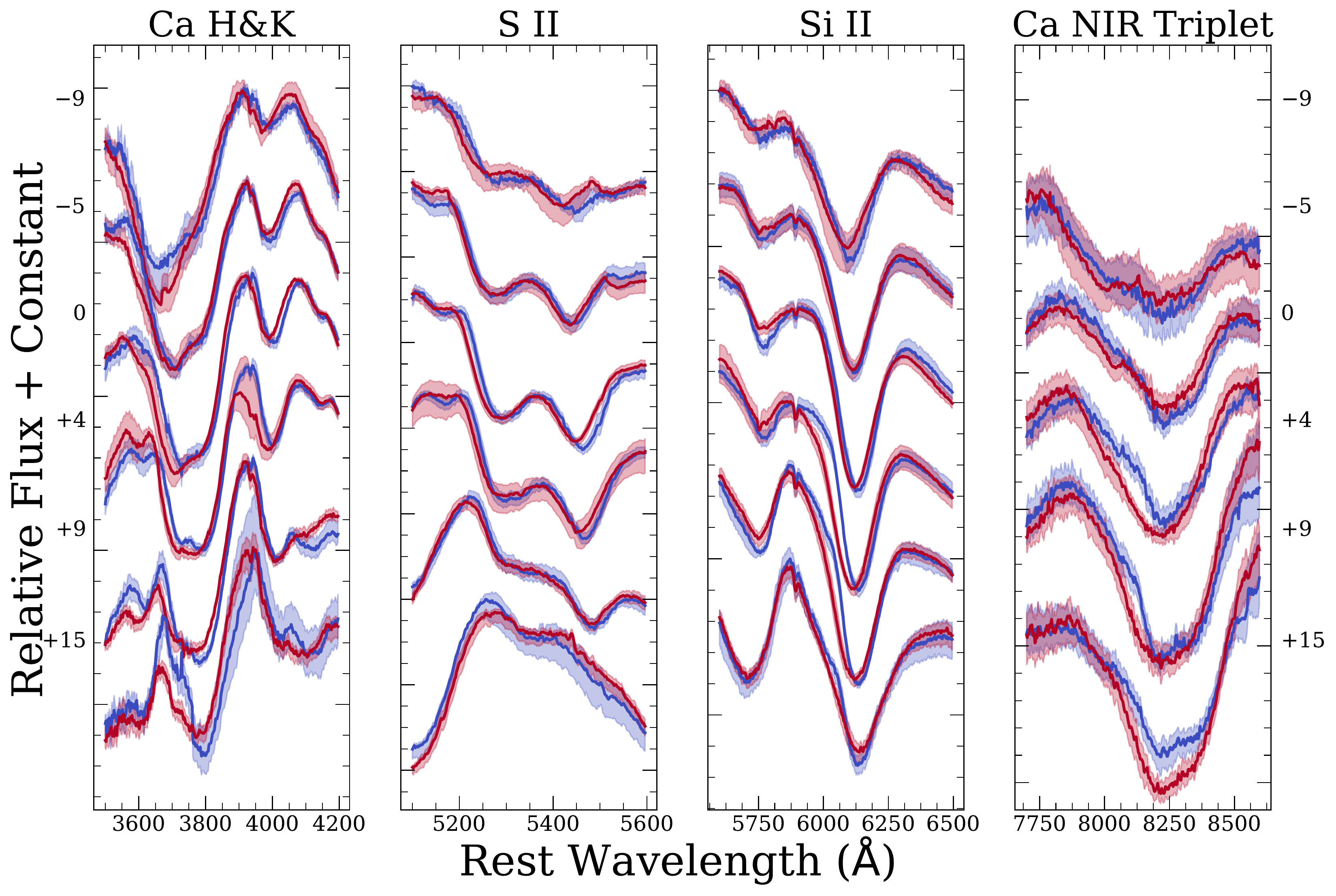}
\caption{Time series of composite spectra created from subsamples of SNe with positive (blue) and (negative) HRs focusing on the Ca H\&K (first panel), \ion{S}{II} (second panel), \ion{Si}{II} (third panel), and \ion{Ca}{II} NIR triplet (fourth panel) features. Blue and red curves/shaded regions correspond to the positive-HR and negative-HR composite spectra respectively. The effective phase from top to bottom of the sets of composite spectra are $-9$ (excluding the \ion{Ca}{II} NIR triplet), $-5$, 0, +4, +9, and +15~days.}\label{fig:zoom}
\end{center}
\end{figure*}

In \autoref{fig:zoom} we display the spectral evolution of the Ca H\&K, \ion{Si}{II} $\lambda$4130, \ion{S}{II} ``W" feature, \ion{Si}{II} $\lambda$5972, $\lambda$6355, and the Ca NIR Triplet in the same 6 sets of HR-binned composite spectra. We do not show composite spectra of the Ca NIR triplet at the earliest epoch because there are not enough SNe in the \code{kaepora} HR sample at that epoch to get 5 SNe per wavelength bin in that region. The fluxes of the HR-binned composite spectra have been rescaled in each absorption region to best compare the line profiles.  

In the majority of the negative-HR composite spectra (red), the minima of these absorption features appear blueshifted relative to their positive-HR counterparts (blue). With the exception of S~II, which shows a large velocity difference at $-9$~days, the largest discrepancies in velocity occur either in the maximum-light or +4-day composite spectra for each absorption feature. All of the \ion{Si}{II} features ($\lambda$4130, $\lambda$5972, and $\lambda$6355) are more blueshifted in the negative-HR composite spectra at every epoch except +15~days where they have similar velocities. Ca H\&K appears more blueshifted in the negative-HR composite spectra at every epoch. However, the Ca NIR triplet is more variable. The early-phase ($-5$ and 0~days) composite spectra of this feature are very similar, but the later-phase composite spectra show the same velocity trend as the other spectral absorption features. 

We further examined color-corrected flux ratios other authors found to correlate with HRs.  For the highest-significance ratios, we find differences in the +4-day composite spectra.  Notably, the two wavelengths used for a flux ratio (e.g., 4260 and 4610~\AA\ for $\mathcal{R}^{c}(4610/4260)$; \citealt{blondin11}) are at the edges of a spectral feature (in the example, \ion{Mg}{II}).  For these cases, the flux ratio changes significantly with velocity since the edge of a feature will shift, adjusting the flux at a particular wavelength.  We do not find evidence for different line strengths beyond the velocity difference for these measurements.

\subsection{Hubble Residuals and Color Curves}\label{subsec:color}

\begin{figure*}
    \includegraphics[width=3.2in]{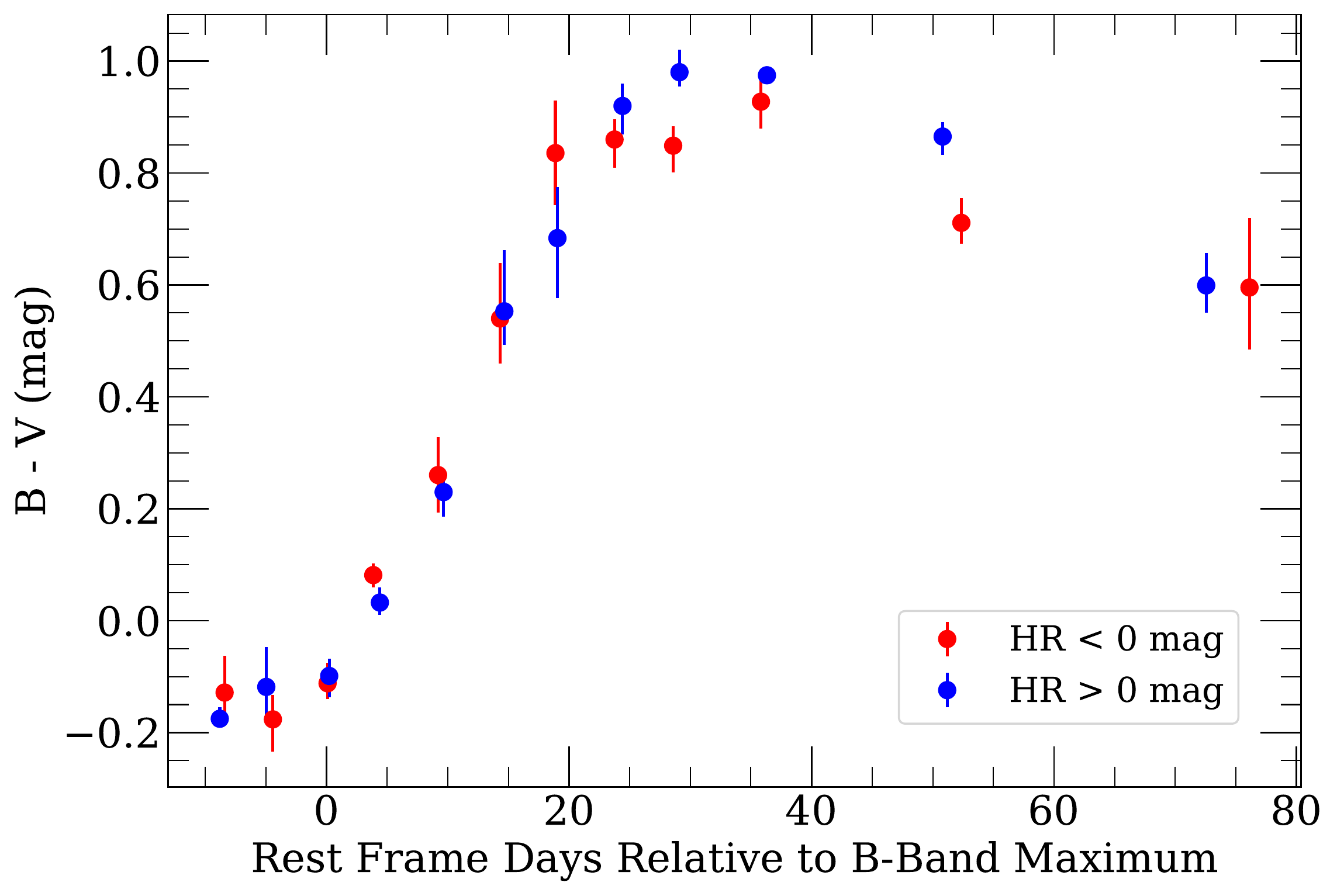}
    \includegraphics[width=3.2in]{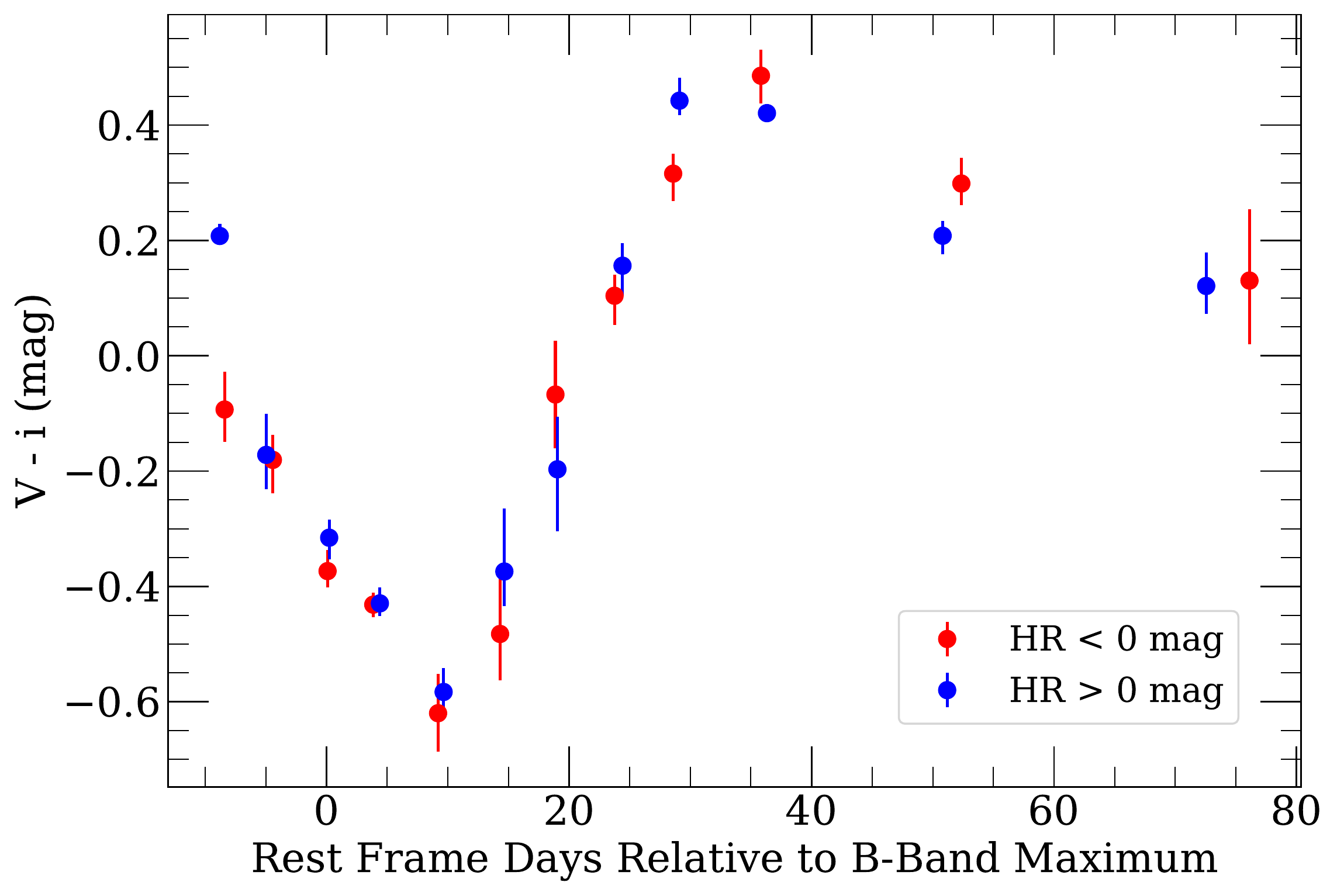}
  \caption{$B-V$ (left) and $V-i$ (right) color curves of our positive- (blue) and negative (red) composite spectra. }\label{fig:color_curves}
\end{figure*}

Despite the velocity differences observed in the spectral absorption features, the continua are similar.  We examine this in more detail by synthesizing $B-V$ and $V-i$ color curves, which we present in \autoref{fig:color_curves}.

The color curves of the HR-binned composite spectra look very similar. There is some tentative evidence that the positive-HR sample is redder than that of the negative-HR sample at $\tau \gtrsim 30$~days, while the $V-i$ color of positive-HR sample is bluer. This potential trend is discussed further in Section \ref{sec:disc}.

\subsection{Hubble Residuals and Absorption Strength}

\begin{figure*}
    \includegraphics[width=6.1in]{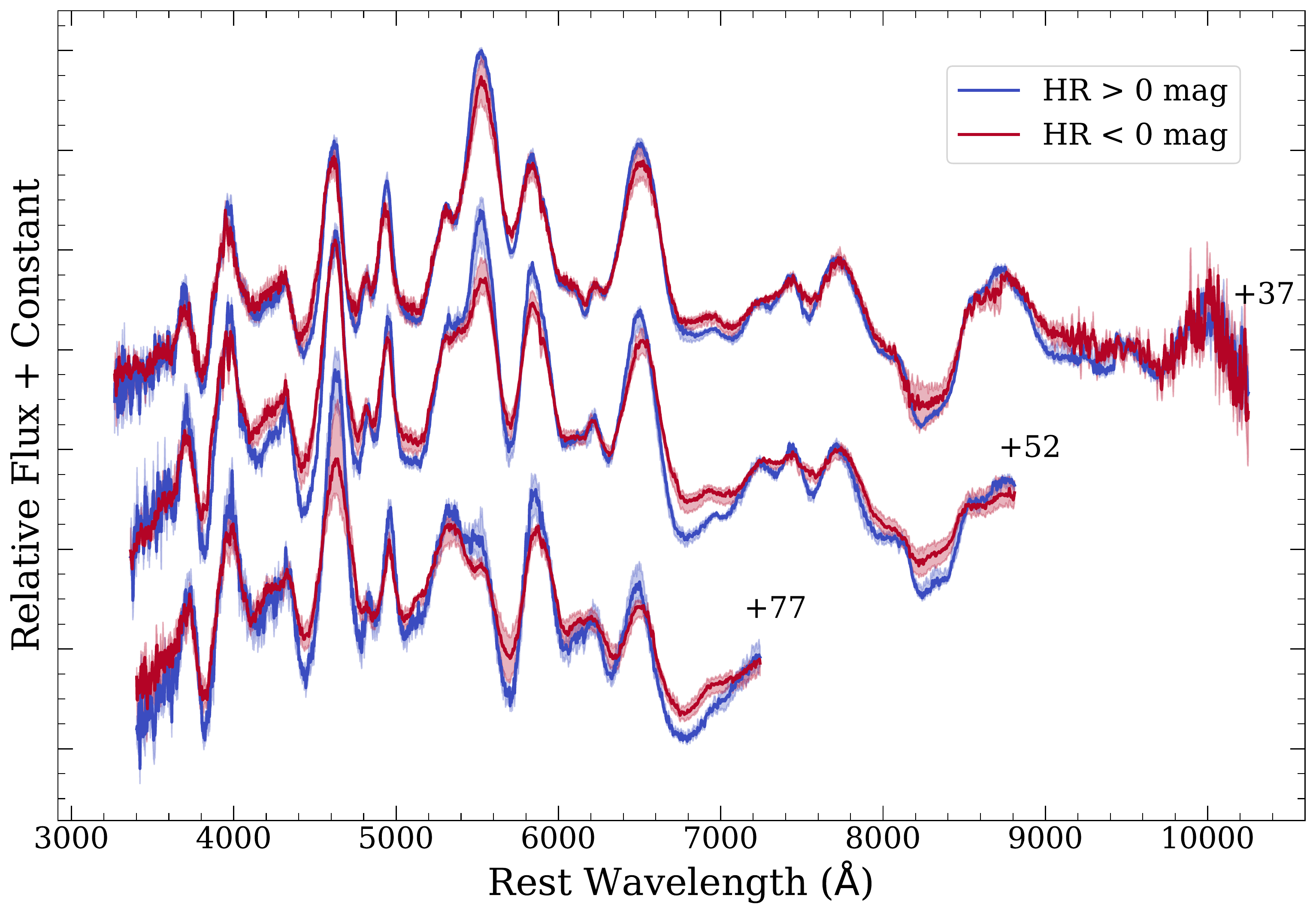}
  \caption{HR-binned composite spectra Composite spectra with effective phases of +37, +52, and +77~days from top to bottom. Blue and red curves/shaded regions correspond to the positive- and negative-HR composite spectra, respectively.}\label{fig:late_phases}
\end{figure*}

In addition to the velocity-HR relationship discussed above, we also see an indication of spectral deviations between HR-binned samples at  phases 1--3 months after peak.  In \autoref{fig:late_phases} we present three sets of HR-binned composite spectra with effective phases of +37, +52 and +77~days from top to bottom, respectively.

At +37~days, the positive-HR and negative-HR composite spectra look very similar. While there are some subtle differences, most of these deviations are contained within the 1-$\sigma$ bootstrap resampling uncertainty regions. However, as the composite spectra progress to later phases, the differences become more significant.

In the three displayed spectra, the negative-HR composite spectrum has weaker features than the positive-HR composite spectrum. This relationship is most obvious in the +52-day HR-binned composite spectra, but is also significant in the +77-day HR-binned composite spectra, but with a limited wavelength range and fewer SNe contributing. For this reason, we chose to further examine this trend by examining the +52-day composite spectra in more detail. 

\begin{figure}
    \includegraphics[width=3.2in]{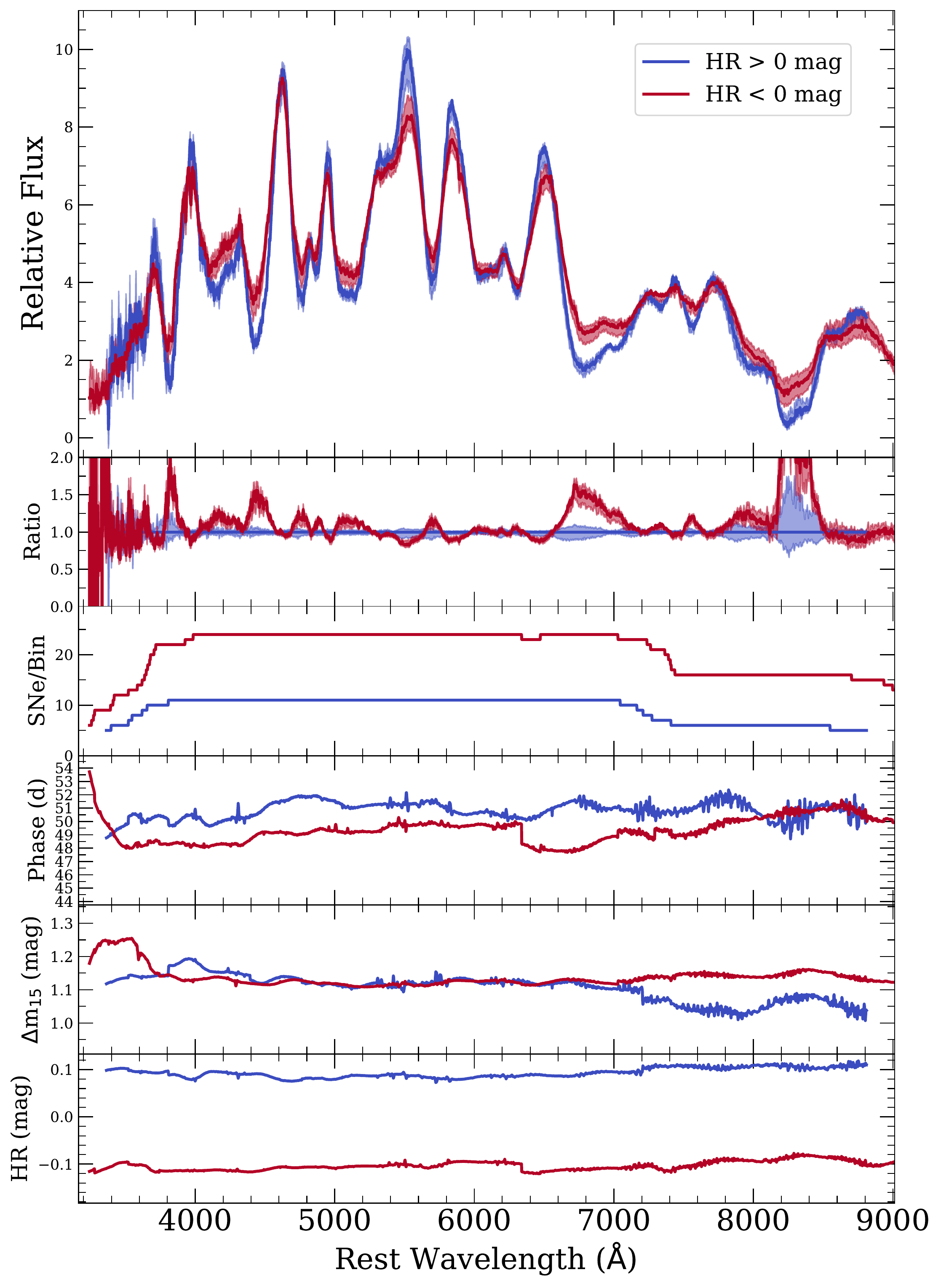}
  \caption{Same as \autoref{fig:p4_panels} but comparing the +52-day HR-binned composite spectra.}\label{fig:p52_strength}
\end{figure}

\autoref{fig:p52_strength} is the same format as \autoref{fig:p4_panels} and shows these composite spectra along with some more detailed, wavelength-dependent information. These composite spectra were constructed using a phase bin of +42 -- +62~days and \dm\ bins of 0.7 -- 1.8~mag. The effective phase, \dm, and HR of the positive-HR composite spectrum are 50.8~days, 1.11~mag, and 0.09~mag, respectively (52.4~days, 1.08~mag, and $-0.11$~mag, for the negative-HR composite spectrum, respectively). While the phase bin is large, the average phase only differs by more than 3~days ($<$5\% of the time since explosion) in 6\% of all wavelength bins. Similarly for \dm\ (fifth panel), the average \dm\ at every wavelength only differs by more than 0.1~mag in $14\%$ of wavelength bins. It is also noteworthy that the positive-HR and negative-HR sample sizes are very different in this phase range. With the positive-HR sample containing 18 spectra of 11 SNe, and the negative-HR sample containing 60 spectra of 31 SNe. While these spectral differences seen at these later epochs are intriguing, the small number of contributing SNe limits our inference.

A difference like this could be attributed to a difference in the overall continuum level of SNe with different HRs. The fractional depth of spectral features in SNe~Ia with a larger overall flux levels should be smaller. \citet{wang19} has suggested that a $B$-band offset in HV SNe at phases $>$+40~days could be due to a light echo from circumstellar dust. A light echo could dilute the specral features.  However, our composite spectra do not exhibit a large color difference at this phase (Section~\ref{subsec:color}), disfavoring a light-echo explanation for the difference in spectral feature strength.

It is also possible that host-galaxy light contamination is causing the observed feature strength differences. This could also change the continuum level, diluting the strength of features. Since the difference between the spectra does not look like a galaxy spectrum, this scenario is less likely.
%The only way to test this effect would be to look at host-galaxy surface brightness at the SN positions. Regardless,
Since there are only 11 SNe in the positive-HR sample, we caution interpretation of this difference until a larger sample is obtained. 

% We attempted to quantify this difference by developing a metric for overall feature strength in a spectrum. First, we define a pseudo-continuum by smoothing the spectrum on a large velocity scale ($d\lambda / \lambda$ = 0.1 in \citetalias{siebert19}). We then take the absolute sum of the difference between the spectrum and pseudo-continuum and divide by the total wavelength range. 

% In \autoref{fig:p52_strength}, we present measurements of this feature strength proxy for our HR-binned composite spectra (blue points), and the individual SNe contributing to our HR-binned composite spectra (orange points). The measurements of this metric from the composite spectra seem to represent to the underlying spectra well. While there is not a tight correlation between feature strength proxy and HR, there is a potential trend. In comparison to the negative-HR bin, there does appear to be a lack of SNe with feature strength proxies smaller than $\sim 0.12$ in the positive-HR bin. 

% \begin{figure}
%     \includegraphics[width=3.2in]{hr_late_features.pdf}
%   \caption{Individual feature strength proxies versus HR of the SNe contributing to our +52 day composite spectra in \autoref{fig:p52_strength}. Orange points correspond to individual SNe, and the blue points correspond to the measurements from our composite spectra.}\label{fig:p52_strength}
% \end{figure}

\section{Discussion}\label{sec:disc}
Composite spectra provide an agnostic approach to understanding spectral correlations with HR. Through this approach we find that velocity is the most important optical spectroscopic indicator of HR after all typical corrections are made. Quantifying the significance of the velocity-HR relationship is challenging since we observe this trend in several spectral absorption features and at several epochs. Most methods yield a statistically significant result.

We find that SNe with negative HRs tend to have higher ejecta velocity.  Similarly, we find that SNe with higher ejecta velocity tend to have negative HRs after all light-curve and host-galaxy corrections are applied.  That is, using modern techniques as done for cosmological analyses, SNe~Ia with high (low) ejecta velocity have measured distance that are biased low (high).

In this section, we further discuss the significance of our results, and the implications a velocity distance bias would have on SN~Ia cosmology results.

\subsection{Temporal Velocity Evolution and HR Differences}
\begin{figure}
    \includegraphics[width=3.2in]{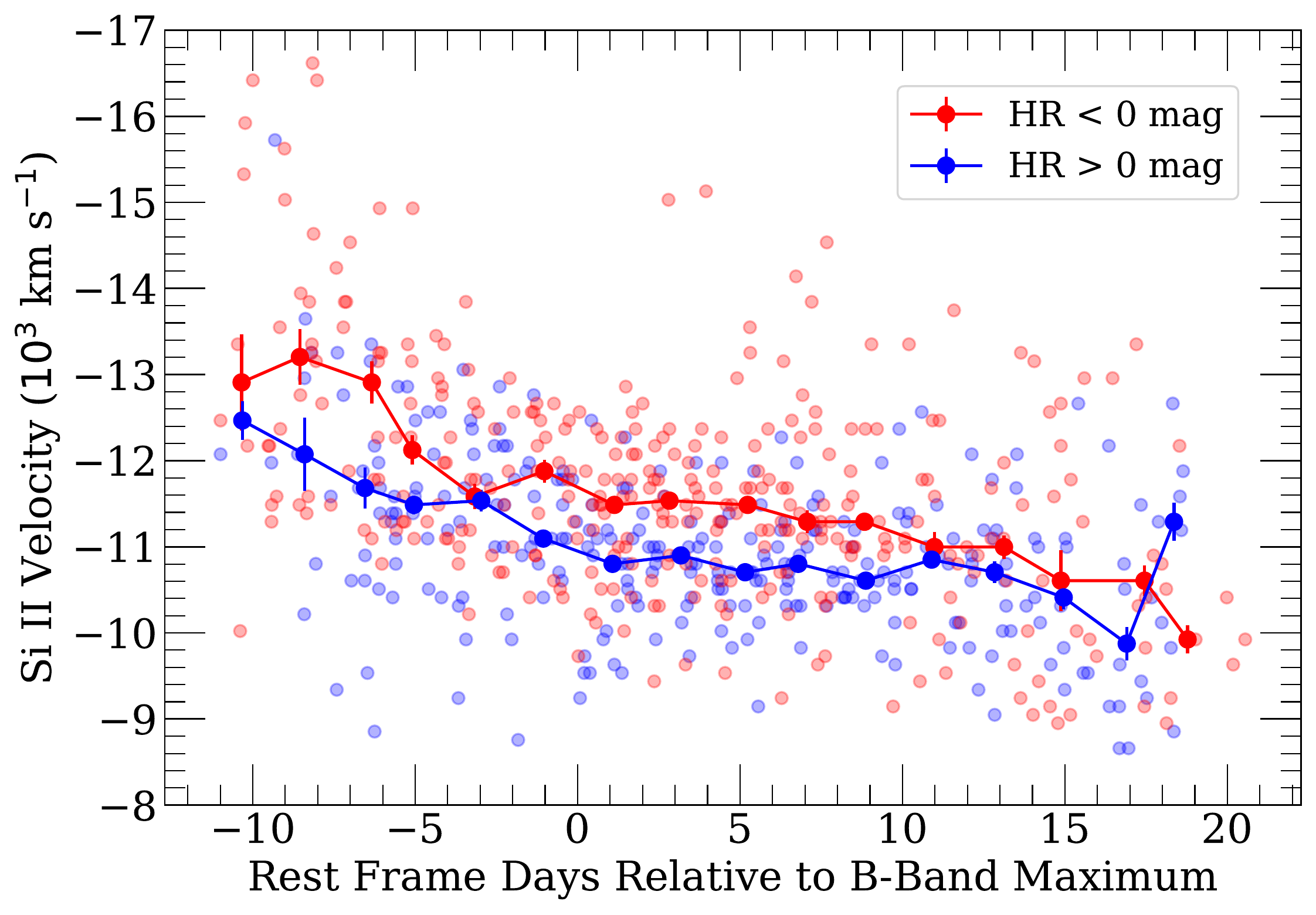}
  \caption{\ion{Si}{II} velocity evolution for the HR sample. Individual points are measurements made from individual spectra. Solid connected points are binned medians using the bin sizes of 2~days and the error bars are the median absolute differences, dividing by the square-root of the number of points in a bin, for each bin. Blue and red points correspond to measurements from SNe in the positive-HR and negative-HR samples, respectively. We estimate that the negative-HR sample has a mean velocity that is at least $500$ km s$^{-1}$ larger than the negative-HR sample at the $2.8\sigma$ level.}\label{fig:v_all_spec}
\end{figure}

In \autoref{fig:v_all_spec}, we further examine the evolution of $v_{\mathrm{Si\; II}}$. The individual blue (red) points correspond to \ion{Si}{II} $\lambda$6355 velocity measurements from individual spectra whose SNe have positive-HR (negative-HR). The solid blue and red points connected by lines are the 2-day binned medians of these velocity measurements, and the error bars are the median absolute difference within each bin.  While the scatter is large, it is clear that the negative-HR sample has consistently higher velocities on average than the positive-HR sample until $\sim$+15~days after maximum light. Over all bins in this phase range, the negative-HR sample is on average 500~km~s$^{-1}$ above the negative-HR sample, consistent with the +4-day subsample. 

However, it is also important to note that these median measurements are correlated. While the phase bins do not overlap and individual measurements contribute to only a single median value, the same SN may have multiple spectra that cover a range of phases. We quantify this correlation with a simple Monte Carlo simulation. The positive-HR and negative-HR samples displayed in \autoref{fig:v_all_spec} contain data from 52 and 62 SNe, respectively. From the full sample, we randomly generate two samples of SNe with these sample sizes. Each SN maintains all of its original velocity measurements. In our original measurements, the negative-HR sample has a higher median velocity in 14 out of 16 phase bins and has a median ejecta velocity higher by at least 500~km~s$^{-1}$ in 9 out of 16 phase bins. We find that the mock negative-HR sample produces a similar distribution of velocity residuals by chance in only 0.6\% of trials.

\subsection{Impact of SALT2 Corrections}

\begin{figure}
    \includegraphics[width=3.2in]{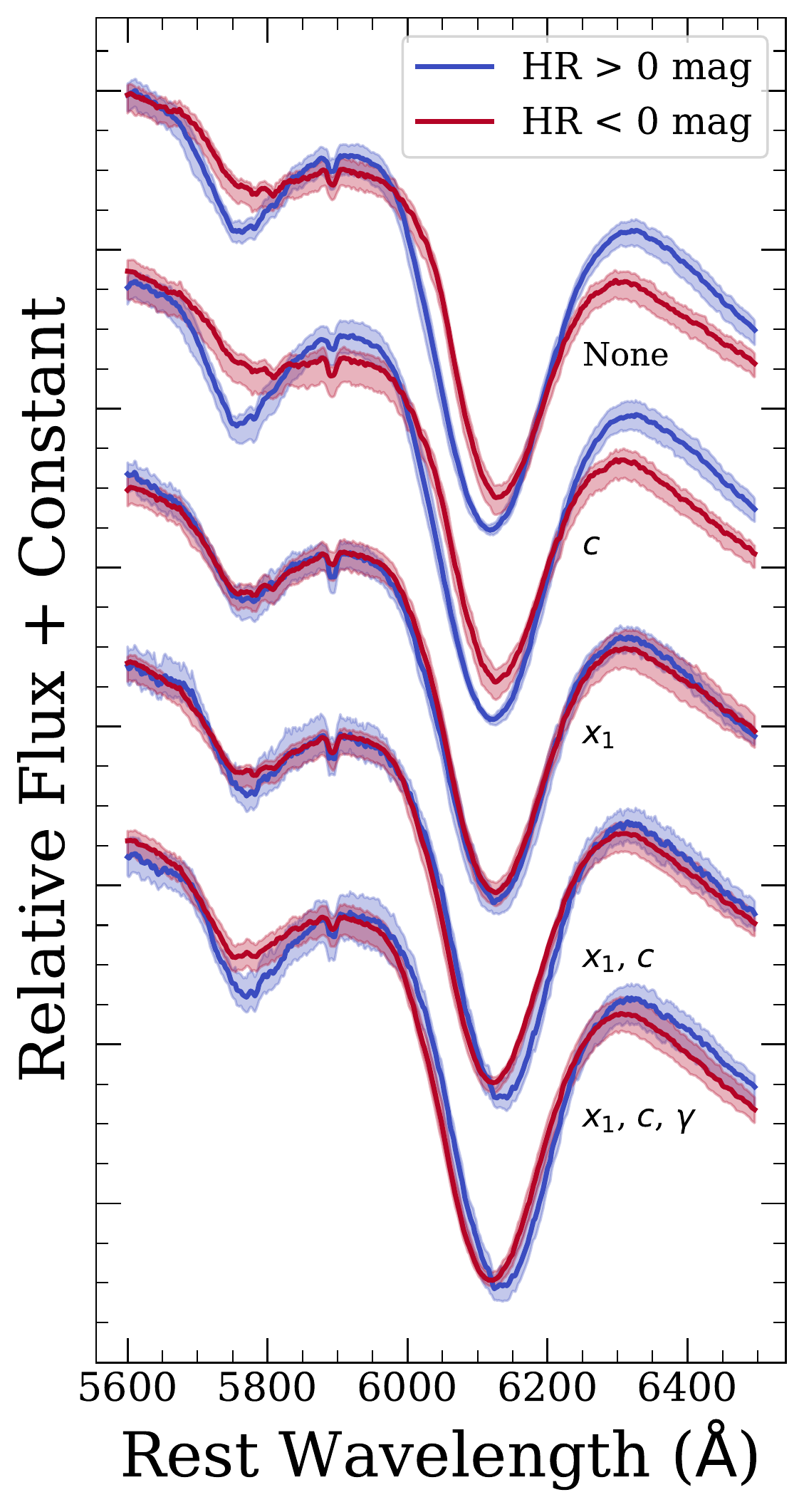}
  \caption{Same as \autoref{fig:hr_corr} but zoomed in on the \ion{Si}{II} $\lambda$6355 feature.}\label{fig:hr_corr_si}
\end{figure}

We examine the impact of the velocity difference on distance modulus biases and cosmological analyses. In \autoref{fig:hr_corr_si} we present the maximum-light HR-binned composite spectra from \autoref{fig:hr_corr} focusing on the \ion{Si}{II} $\lambda$6355 feature. As in \autoref{fig:hr_corr}, the top to bottom sets of positive-HR (blue) and negative-HR (red) composite spectra have increasing corrections applied before calculating HRs. 

We find that for both the $x_{1}$ and $c$-corrected and $x_{1}$, $c$, and $\gamma$-corrected composite spectra, the negative-HR composite spectra clearly have a higher $v_{\mathrm{Si\; II}}$ than positive-HR composite spectra. This means that the effect is present regardless of if a host-mass step ($\gamma$) correction is applied. Additionally, we are not introducing a velocity-HR correlation when accounting for the host-mass step. When we only apply the $x_1$ correction (third set of composite spectra), the velocity difference between the HR bins disappears indicating that, if there is still a velocity effect, it is not a dominant influence on HRs.

\subsection{Consistency with Predictions from FSK11}

The trend between HR and velocity has important implications for cosmological analyses. Here we quantify the improvement of precision obtained by applying the velocity step presented in \autoref{fig:hrstep}.

The intrinsic HR scatter for a sample is defined as the additional uncertainty required to achieve $\chi_{\nu}^2 = 1$. The intrinsic scatter in the +4-day sample improves from 0.094~mag to 0.082~mag after applying the velocity step. This improves the distance precision of each SN by 20\%. However, this result is not significant ($0.5\sigma$). While potentially intriguing, this result is also still consistent with \citet{blondin11} and \citet{silverman12}, who found a $<$10\% improvement in the HR weighted root-mean square (WRMS) when accounting for SN~Ia maximum-light velocities. Despite the small improvement, \citet{blondin11} found evidence for a weak correlation between $v_{Si\; \mathrm{II}} ^0$ and $x_1$/$c$-corrected HRs (absolute Pearson correlation coefficient of 0.40).

\citetalias{fsk11} noted a correlation between velocity and color (VCR), which could have implications for a HR bias. \citetalias{fsk11} measured an intrinsic color by comparing the color of a SN to the expected color for a SN with a similar light-curve shape corrected peak brightness that is only reddened and dimmed by dust.  Deviations in color from the expected relationship were interpreted as intrinsic.

\citetalias{fsk11} found a linear correlation between \ion{Si}{II} velocity at maximum light and $(B_{\mathrm{max}} - V_{\mathrm{max}})_0$, the intrinsic difference between maximum $B$- and $V$-band brightnesses (the velocity-color relationship; VCR).  Specifically, they found
\begin{equation}
\begin{split}
(B_{\mathrm{max}} - V_{\mathrm{max}})_0 = (-0.39 \pm 0.04) - (0.033 \pm 0.004) \\
\times \left ( v_{\mathrm{Si\; II}} ^0/1000\; \mathrm{km~s^{-1}} \right ).
\end{split}
\label{eq:int_color}
\end{equation}
Since the SALT2 $c$ parameter behaves in a similar way to $B-V$ color \citet{guy05}, the VCR should produce, assuming SALT2 does not somehow account for this affect already, a step between low- and high-velocity SNe of
\begin{equation}
\mathrm{HR}_{\rm low} - \mathrm{HR}_{\rm high} \approx (0.033 \pm 0.004) \times \frac{v_{\rm low} - v_{\rm high}}{1000\; \mathrm{km~s^{-1}}} \times \beta,
\label{eq:int_color_2}
\end{equation}
where $\beta = 3.15$ and ``low'' and ``high'' subscripts correspond to average measurements from the $v_{\mathrm{Si\; II}} > -11{,}000$~km~s$^{-1}$ and $v_{\mathrm{Si\; II}} < -11{,}000$~km~s$^{-1}$ subsamples, respectively.

The prediction that higher-velocity SNe should have negative HRs is a direct prediction from the direction of the correlation between velocity and color and is independent of our measurements.  However, we now use mean velocity measurements of the low- and high-velocity +4-day HR samples ( $-10{,}510$ and $-11{,}880$~km~s$^{-1}$, respectively) to estimate that VCR should produce a HR step of $0.14 \pm 0.04$~mag for our sample. This step size is consistent with the HR step that we observe ($0.091\pm 0.025$~mag) and has an important physical interpretation.  Higher-velocity SNe~Ia, which have redder intrinsic continua, are over-corrected for their color, making them appear closer away than they are.  Notably, this would lead to corrected magnitudes that are brighter (negative-HR) than lower-velocity SNe~Ia. Since the distribution of $c$ values within our high- and low-velocity subsamples are very similar, we suggest that $c$ does not fully account for intrinsic color differences due to velocity. The reduction in intrinsic scatter that we measure could indicate that chromatic scatter could account for $\sim$20\% of the observed intrinsic scatter.

% The \citetalias{fsk11} relation is derived from maximum-light measurements, and thus $v_{\rm low}$ and $v_{\rm high}$ in \autoref{eq:int_color_2} should also formally be maximum-light measurements.  While our estimate above used velocities from 4~days after peak brightness, the {\it difference} in velocity measured at +4~days should be similar to the velocity difference at maximum light. 

While our measurement is consistent with the prediction of \citetalias{fsk11}, there is significant overlap between the \citetalias{fsk11} sample and the HR sample presented in this work.  Although \citetalias{fsk11} did not examine HRs, it should not be surprising that we observed a similar HR offset {\it if} that offset is caused by VCR.

Given the VCR presented in \citet{fsk11}, we might expect the colors of our HR-binned composite spectra to be different. However, we do not see significant differences in the $B-V$ and $V-i$ colors curves presented in \autoref{fig:color_curves}. This may not be inconsistent since the continua are corrected using a reddening estimate from MLCS \citep{jha07}, while the HR is determined from SALT2 \citet{Guy10}.  It is possible that MLCS, which attempts to separate intrinsic color and dust reddening is less affected by the VCR.

\citetalias{fk11} found that when using the rudimentary corrections of \dm\ and peak color, accounting for the VCR improved the HR scatter from 0.190 to 0.130~mag. By using velocity, our results do not rule out an improvement of this size.  Again, this consistency is not unexpected since the samples in \citetalias{fk11} and this work are not completely independent. % If we use the velocity-color relationship from FSK11, we can attribute a velocity difference of $1{,}000$ km s$^{-1}$ to a difference in $B_{\mathrm{max}} - V_{\mathrm{max}}$ color of 0.033 mag. The difference cannot be explained by a difference in $c$. The positive-HR, negative-HR, $v > -11{,}000$ km s$^{-1}$, and $v < -11{,}000$ km s$^{-1}$ samples all have average $c$ distributions that are consistent with 0 and not significantly different from each other. If SNe~Ia with higher velocities do have redder intrinsic continua, then it is possible that cosmological distance estimators are over-correcting these events for reddening caused by dust. This would lead to corrected magnitudes that are overluminous (negative-HR) than lower velocity SNe~Ia. Using the Tripp formula \citep{tripp98} and $R_V = 2.5$, we estimate that $B_{max} - V_{max}$ color difference of 0.033 mag could propagate to a difference in HR = 0.08 mag. This is also consistent with the HR step that we measure when subdividing the sample by velocity. We typically estimate intrinsic scatter to be $\sim 0.08-0.14$ mag (\mrs{CITE}). If velocity differences that we observe can be attributed to the velocity-color relationship, then our results would suggest that chromatic scatter could account for $33-100\%$ of the observed intrinsic scatter. 

\section{Conclusions}\label{sec:conc}

We have used the open-source relational database \code{kaepora} to generate a variety of composite spectra with different average properties in order to investigate potential spectral variation with HRs. Using sets of composite spectra, we further examine how SNe~Ia with different HRs differ spectroscopically.  Our main results can be summarized as follows:

\begin{enumerate}
  \item There are several spectral differences between SNe~Ia with different HRs.  This indicates both that current distance estimators that rely on photometric and host-galaxy measurements alone are not capturing the full physical diversity of SNe~Ia and that measuring these spectral differences provides the possibility of improving distance measurements.
  \item There exists a trend between \ion{Si}{II} velocity and HR. Using a sample of 62 SNe~Ia, we find that SNe with negative-HRs tend to have higher \ion{Si}{II} $\lambda$6355 velocities. We measure a HR-\ion{Si}{II} $\lambda$6355 velocity step at  $-11,000$ km s$^{-1}$ of $0.091\pm 0.035$~mag. This step is consistent with a VCR \citep{fsk11}, and correcting for the velocity step may improve distance precision by 20\%.
  \item  Using all individual spectra ranging from $-13$ to +22~days, we find $2.8\sigma$ evidence for a \ion{Si}{II} $\lambda$6355 velocity difference $>$500~km~s$^{-1}$ between the positive-HR and negative-HR samples at all epochs.
  \item A similar velocity offset between positive-HR and negative-HR samples is observed at a variety of epochs and in multiple spectral features. For negative-HR composite spectra we observe larger blueshifts in Ca H\&K, \ion{Si}{II} $\lambda$4130, \ion{Si}{II} $\lambda$5972 and $\lambda$6355, and the Ca II NIR triplet, and at phases of $-9$ to +15~days. The differences are apparent across the line feature, with the negative-HR spectra having broader features with bluer blue edges.  In many different ways across phase, atomic species, and line profile, the negative-HR SNe have indications of higher ejecta velocity, making the above results more significant. Additionally, we use near-maximum light spectra to estimate $v^0_{\mathrm{Si\; II}}$ for a sample of 115 SNe~Ia. With these measurements we estimate a $v^0_{\mathrm{Si\; II}}$-HR step of $0.068 \pm 0.027$~mag when sample intrinsic scatter is included in HR uncertainties.
  \item Using our maximum-light composite spectra, we find that this velocity difference is apparent for HRs derived with and without accounting for the host-galaxy mass step. This indicates that the velocity difference is not induced by making this distance modulus correction, and the host-mass step is not caused by velocity. 
  \item At late epochs ($+37-77$ days) we observe that the negative-HR sample produce composite spectra that appear to have overall weaker spectral features than the positive-HR sample. This difference appears to strengthen with time, however, more spectra are needed validate this effect.
  \item All results are consistent with a VCR \citepalias{fsk11}.  In this scenario, high-velocity SNe~Ia have intrinsically redder continua than low-velocity SNe.  The redder continua is over-corrected using current distance estimators, causing a distance bias.
\end{enumerate}

A velocity-HR step of $\sim$0.1~mag has important implications for cosmological analyses. Currently there is not significant evidence that the velocity distributions of SNe~Ia evolve with redshift \citep{blondin06, Foley12:highz}. However if the average velocity changes with redshift, we would introduce a systematic bias in our cosmological parameter estimates. We should gather more high-redshift spectra to further explore this effect. Specifically, we have presented evidence that the velocity-HR trend manifests as a velocity difference in numerous absorption features. These potential differences should be further explored in both low- and high-redshift samples. Also, \citet{sugar} recently developed SUGAR to improve the spectral description of SNe~Ia. Using a ``PCA-like'' method, they showed that the second-most important factor was strongly correlated with \ion{Si}{II} and \ion{S}{II} velocity. The variation in this factor was also related to variation in color, and was not correlated with the SALT2 $x_1$ or $c$ parameters. This supports our conclusion that SALT2 does not fully capture spectral variations associated with ejecta velocity.

Since ejecta velocity trends with HR, we encourage more spectral observations of SNe~Ia, perhaps especially immediately after maximum light (+2 -- +7~days) to provide the best unbiased distance measurements and to increase sample sizes. Since the VCR is consistent with our observations, it might be possible to glean additional distance information from color curves without directly measuring velocities. However with our current sample and techniques, we do not see significant differences in the $B-V$ or $V-i$ color curves of our HR-binned composite spectra at a variety of epochs. This indicates that spectra are likely needed to fully capture the velocity effect. While the spectral differences at later phases should be further explored, it is less feasible to get spectroscopic follow-up observations of SNe~Ia in a cosmological sample at these late epochs. 

Future cosmological experiments such as the Large Synoptic Survey Telescope (LSST) and the {\it Wide-Field InfraRed Space Telescope} ({\it WFIRST}) will discover and photometrically follow large samples of SNe~Ia ($10^{5}$ -- $10^{6}$), but only a small fraction will have spectroscopy \citep{Hounsell18, LSST18}.  Careful choices must be made, especially now that \textit{WFIRST}'s spectroscopic capabilities are limited to a slitless prism and grisms, to fully sample the velocity distribution and to determine if that distribution changes with redshift. The velocity-HR effect must be analyzed in detail to properly plan for and leverage these experiments.

\section*{Acknowledgements}
M.R.S.\ is supported by the National Science Foundation Graduate Research Fellowship Program Under Grant No.\ 1842400.  D.O.J.\ is supported by a Gordon \& Betty Moore Foundation postdoctoral fellowship at the University of California, Santa Cruz. The UCSC team is supported in part by NASA grant NNG17PX03C; NSF grants AST--1518052 and AST--1815935; the Gordon \& Betty Moore Foundation; the Heising-Simons Foundation; and by a fellowship from the David and Lucile Packard Foundation to R.J.F.

\bibliographystyle{mnras}
\bibliography{ref}

% Don't change these lines
\bsp	% typesetting comment
\label{lastpage}
\end{document}